\documentclass[%
 reprint,
superscriptaddress,
showpacs,preprintnumbers,
 amsmath,amssymb,
 aps,
pra,
]{revtex4-1}
\usepackage[dvipsnames]{xcolor}
\usepackage[colorlinks=true,bookmarks=false,linkcolor=NavyBlue,urlcolor=NavyBlue,citecolor=NavyBlue,breaklinks]{hyperref}
\usepackage{amsmath}
\usepackage{amsfonts}
\usepackage{graphicx}
\usepackage{color}
\usepackage{braket}
\usepackage{graphicx}
\usepackage{dcolumn}
\usepackage{bm}

\newcommand{\urow}{\uparrow}
\newcommand{\drow}{\downarrow}

\begin{document}

\title{Tunable spin and orbital polarization in $\mathrm{SrTiO_3}$-based heterostructures}

\author{Cong Son Ho}
 \affiliation{
Department of Electrical and Computer Engineering, National University of
Singapore, 4 Engineering Drive 3, Singapore 117576.}
\author{Weilong Kong}
\affiliation{Department of Physics,
Faculty of Science
2 Science Drive 3
Blk S12, Level 2
Singapore 117551}
\author{Ming Yang}
\affiliation{Institute of Materials Research and Engineering,  
2 Fusionopolis Way, Innovis, Singapore 138634.}
\author{Andrivo Rusydi}
\affiliation{Department of Physics,
Faculty of Science
2 Science Drive 3
Blk S12, Level 2
Singapore 117551}
\author{Mansoor B. A. Jalil}%
 \email{elembaj@nus.edu.sg}
\affiliation{
Department of Electrical and Computer Engineering, National University of
Singapore, 4 Engineering Drive 3, Singapore 117576.}


\begin{abstract}
We formulate the effective Hamiltonian of Rashba spin-orbit coupling (RSOC) in $\mathrm{LaAlO_3/SrTiO_3}$ (LAO/STO) heterostructures. We derive analytical expressions of properties, e.g., Rashba parameter, effective mass, band edge energy and orbital occupancy, as functions of material and tunable heterostructure parameters. While linear RSOC is dominant around the $\Gamma$-point, cubic RSOC becomes significant at the higher-energy anti-crossing region. We find that linear RSOC stems from the structural inversion asymmetry (SIA), while the cubic term is induced by both SIA and bulk asymmetry. Furthermore, the SOC strength shows a striking dependence on the tunable heterostructure parameters such as STO thickness and the interfacial electric field which is ascribed to the quantum confinement effect near the LAO/STO interface. The calculated values of the linear and cubic RSOC are in agreement with previous experimental results. 

\end{abstract}

\maketitle



\section{Introduction}
In the search for a novel platform for spintronics devices, oxides heterostructures has emerged as one of the most promising candidates due to their excellent electrical and magnetic properties \cite{Andrivo:2018}. The formation of two-dimensional electron gas \cite{ohtomo2004high,2DEG1,2DEG2,2DEG3} and magnetism \cite{brinkman2007magnetic,kalisky2012critical,li2011coexistence,dikin2011coexistence, Yangming:2018} at the interface of transition metal oxides such as $\mathrm{LaAlO_3/SrTiO_3}$ (LAO/STO) gives rise to a unique opportunity to study and implement spintronics in a single platform \cite{hwang2012emergent}. 

A key requirement to spintronics application is the presence of spin-orbit coupling (SOC). As the inversion symmetry at the interface of the two transition metal oxides is inherently broken, there exists SOC of Rashba type \cite{RN338,RN339,RN337}. Magnetotransport measurements have detected  a substantial spin-orbit splitting of 2-10 meV at the LAO/STO interface \cite{RN337,RN406,RN407,RN372}, while a much stronger spin-splitting of 90 meV has been observed on the surface of STO \cite{RN415}. Strikingly, the spin splitting is not just of $k$-linear \cite{RN337} but also $k$-cubed Rashba types \cite{RN408}. Furthermore, the interfacial Rashba coupling is versatile, and can be tunable via electrical gate \cite{RN337,RN338,RN339}, and via crystal orientation selection \cite{RN365,RN369,RN370}.  For comparison, the above splitting is higher than that of 1-5 meV in typical semiconductor heterostructures \cite{winkler2003spin}. The strong Rashba coupling in LAO/STO has led to  the realization of various spintronics effects such as spin-to-charge conversions \cite{RN412,jin2016nonlocal,RN413,RN414,RN418} and spin-orbit torques\cite{RN411}. The fact that these phenomena are interfacial in nature allows us a much better degree of external control of its electrical and magnetic properties compared to conventional spintronic materials such as ferromagnetic metals \cite{huang2018interface,pai2018physics}. More importantly from the application standpoint, the conducting layer is protected against external perturbations as it is sandwiched between conventional oxides insulators \cite{huang2018interface,pai2018physics}, which makes it more robust in comparison to the exposed surface states of topological insulators. 

At the same time, theoretical models have been proposed to elucidate the rich SOC properties of the STO-based heterostructures, which are not captured by the standard linear Rashba theory \cite{Rashba1984}. Combined first-principles calculations and tight-binding approach \cite{RN339,PhysRevB.88.041302,Zhou:PRB15} revealed a multi-orbital Rashba effects, of the linear and cubic types, at the LAO/STO interface layer. Nevertheless, the crystal field, one of the ingredients of the SOC which accounts for the properties of the hetetrostructure, had been deliberately introduced, and the contribution of the bulk STO as well as the inter-orbital hopping was ignored. Significantly, the large spin splitting at the anti-crossing region, a signature of the multi-orbital effects, has not been fully elucidated. Recently, a theory based on k.p model \cite{RN401,k.p2}, including the contribution of the bulk STO, predicted a spin splitting induced by the inter-orbital hopping and interfacial band-bending. However, a complete picture of the multi-orbital Rashba SOC is still lacking.

In this work, we aim to construct the effective Hamiltonian of Rashba spin-orbit coupling system in STO-based structures, taking into account the underlying factors such as the confinement effect, as well as interface effects including inversion symmetry breaking (ISB) and band-bending.  We derive analytical expressions of properties, e.g., Rashba parameter, effective mass, band edge energy and orbital occupancy, as functions of material and tunable heterostructure parameters. We show that the type and strength of spin-orbit coupling as well as the orbital selectivity are tunable by controlling geometric factor such as STO thickness and by applying electric field via gate voltage. The thickness and gate control of SOC provides a possible avenue for optimization of Rashba SOC for spintronics applications and devices. 

\section{Model}
 We consider a 2DEG system in STO-based heterostructure. The 2DEG can be generated by interfacing STO with either LAO \cite{RN358} or vacuum with $\delta$-doping \cite{RN370}. In the low energy limit, the electron in bulk STO is described by following Hamiltonian
\begin{equation}\label{Hfull}
H_\mathrm{full}=H_{t_{2g}}+H_{\mathrm{ISB}}+H_{\mathrm{SO}}+V(z).
\end{equation}
In the above $H_{t_{2g}}$  is the kinetic energy of the $d$-electron in the  ${\Gamma _{25}^+}$ band in cubic crystal, $H_{\mathrm{ISB}}$ is the inter-orbital hopping, $H_{\mathrm{SO}}$ is the atomic spin-orbit interaction,  and the last term is the electrical confinement potential in the  $z$-direction $V(z)=\mathcal{E}_zz$. 

  In the $t_{2g}$ orbital-spin basis $\Phi_{t_{2g}}=\left\{|yz,\urow\rangle, |zx,\urow\rangle, |xy,\urow\rangle, |yz,\drow\rangle,|zx,\drow\rangle, |xy,\drow\rangle\right\}$, with $yz, zx$ and $xy$ representing the $d_{yz}, d_{zx}$, and $d_{xy}$ orbitals, respectively,  the electron in the bulk Ti  $t_{2g}$ band can be described by the effective $\bf{k\cdot p}$ Hamiltonian \cite{RN400,RN401}
\begin{eqnarray}
H_{t_{2g}}=\mathbb{I}_2 }\otimes h_{\Gamma _{25}^+  ,
\end{eqnarray}
where $\mathbb{I}_2$ is $2\times 2$ unit matrix. The diagonal terms of $h_{\Gamma _{25}^+}$ is given by $\langle yz,\sigma|h_{\Gamma _{25}^+}|yz,\sigma\rangle=L k_x^2+M(k_y^2+k_z^2 ),$ and the off-diagonal terms are $\langle yz,\sigma|h_{\Gamma _{25}^+}|zx,\sigma\rangle=Nk_x k_y$, with effective mass parameters $L, M, N$, and $\sigma=\urow,\drow$. The other elements follow by exchanging the $x,y,$ and $z$ indices. In the above, the off-diagonal elements can be considered as the inter-orbital hopping in the bulk STO. The fit of  the $\bf k\cdot p$ model to the energy dispersion from our DFT calculations yield the mass parameters as $L=0.68\ \mathrm{eV\AA^2}, M=9.23 \ \mathrm{eV\AA^2}$, and $N=1.54  \ \mathrm{eV\AA^2}$ [see Appendix \ref{S1} for the details].

 At the same time, in the presence of interface, the inversion symmetry is broken which leads to an additional Hamiltonian \cite{tetragonal:PRB2012,Kim:PRB13,RN402,RN403,Zhou:PRB15}
\begin{eqnarray}\label{ISB}
H_{ISB}=
\mathbb{I}_2\otimes  \gamma\begin{bmatrix}
0&0&2i k_x\\0&0&2i k_y \\-2i k_x&-2i k_y&0
\end{bmatrix},
\end{eqnarray}
up to the first order in $k$, where $\gamma$ is the inversion symmetry breaking (ISB) parameter. In general, the ISB parameter is position-dependent, having a maximum value at the interface and rapidly decreases across at the deeper layers \cite{RN402,RN403}.

The above Hamiltonian is symmetrical with respect to the $xy, yz,$ and $zx$ orbitals. However, this orbital degeneracy will be lifted in the presence of the atomic SOC. The SOC splits the six-fold degeneracy into a four-fold $J = 3/2$ multiplet
of symmetry $\Gamma _8^+$ and a twofold $J = 1/2$ multiplet of symmetry $\Gamma _7^+$. In the original $t_{2g}$ basis, the SOC is expressed as
\begin{eqnarray}\label{HSO}
H_\mathrm{SO}=\xi_\mathrm{SO}
\begin{bmatrix}
 0&i&0&0&0&-1\\-i&0&0&0&0&i\\0&0&0&1&-i&0\\0&0&1&0&-i&0\\0&0&i&i&0&0\\-1&-i&0&0&0&0
\end{bmatrix},
\end{eqnarray}
where the SO splitting is calculated via DFT to be $\Delta_\mathrm{SO}=3\xi_\mathrm{SO}=29.85$ meV [see Appendix \ref{S1}].

	\section{Effective Hamiltonian}
We now turn to derive the Rashba split band structure of a STO-based heterostructure with 2DEG. For a STO film with finite thickness, we first find the eigen-states of (\ref{Hfull}) at the $\Gamma $ point $H_0=H_\mathrm{full}(k_x=k_y=0)$. Subsequently,  an effective Hamiltonian at other $k$ values can be obtained by projecting the full Hamiltonian into the space of these eigenstates.

To simplify the analytical calculation, we transform the Hamiltonian into the basis $|J,m_j\rangle$ of $\Gamma _8^+$ and $\Gamma _7^+$ multiplets , $\Phi_J=\left\{
|\frac{1}{2},\frac{1}{2}\rangle, |\frac{3}{2},\frac{1}{2} \rangle, 
|\frac{3}{2},\frac{3}{2}\rangle, |\frac{1}{2},-\frac{1}{2}\rangle,
|\frac{3}{2},-\frac{1}{2}\rangle,|\frac{3}{2},-\frac{3}{2}\rangle\right\}$, 
in which the atomic SOC matrix in Eq. (\ref{HSO}) is diagonalized [see Appendix \ref{S2}]. In the above basis, the $\Gamma$-point Hamiltonian is block-diagonalized and it can be decomposed into three subsets as $H_0=h_+\oplus h_0\oplus h_-$, where
\begin{eqnarray}
h_\pm=
\begin{bmatrix}
a&\pm q\\
\pm q&b
\end{bmatrix}, \ \ 
h_0=\begin{bmatrix}
c&0\\
0&c
\end{bmatrix},
\end{eqnarray}
in which $a=\frac{1}{3}(L+2M)k_z^2+2\xi_\mathrm{SO}, b=\frac{1}{3}(2L+M)k_z^2-\xi_\mathrm{SO}$, $c=M k_z^2-\xi_\mathrm{SO}$, and $q=\frac{\sqrt{2}}{3}(M-L) k_z^2$. 

 For STO slab with finite thickness $d$, the electron states are quantized along the $z-$direction, and we can make a substitution $k_z\rightarrow -i\partial_z$. The eigenvectors and subband energies of electron can be found by solving the Schrodinger equation
$
(H_0 +V)\psi(z)=E\psi(z),
$
 in which $\psi(z)$ is six-component eigenvector.  

In solving the above Schrodinger equations in a STO slab with finite thickness $d$, we assume that the electron is confined in a quantum well with open boundaries, i.e., $\psi(z=0,d)=0$. This assumption may be understood by considering the character of the boundaries.  On the one hand, at the LAO/STO interface, the La $t_{2g}$ states in LAO layer lie far above the Ti $t_{2g}$ states in the STO layer \cite{2DEG3}. Therefore,  the penetration of the Ti $t_{2g}$ states into the LAO can be neglected \cite{2DEG3} and one can assume an infinite barrier at the interface for the simplicity. On the other hand, the other surface of the STO slab is open to vacuum, and thus an infinite barrier potential is imposed. Similar boundary conditions are also applied for the $\delta$-doped STO slab, where both STO surfaces are open to vacuum. 

In the framework of variational approach, the solutions are given by
$
|\psi_0^\sigma\rangle=f_0(z) |\frac{3}{2},\sigma\frac{3}{2}\rangle,$ and $
|\psi_\pm^\sigma\rangle=f_\pm(z) \left(\cos{\chi_\pm}|\frac{1}{2},\sigma\frac{1}{2}\rangle+\sigma \sin{\chi_\pm}  |\frac{3}{2},\sigma\frac{1}{2}\rangle\right),
$ 
 where the spatial envelope functions read as 
\begin{eqnarray}
f_\tau(z)&&=C_\tau e^{-\beta_\tau z}\sin{\lambda z}.
\end{eqnarray}
Here, $\lambda=n\frac{\pi}{d}$, $C_\tau$ is the normalization constant, and $\beta_\tau$ is the variational parameter for the minimization of the corresponding eigen-energy $E_\tau=E_\tau^{(0)}+\mathcal{E}_zF(\beta_\tau)$, where $E_\tau^{(0)}$ is the eigenvalue energy solution when $V=0$ given by
 $E_0^{(0)}=M\lambda^2-\xi_\mathrm{SO},
 E_\pm^{(0)}=\frac{1}{2}\left[\xi_\mathrm{SO}+(M+L)\lambda^2\right]
\pm\frac{1}{2}\Delta_I$, in which 
\begin{eqnarray}
\Delta_I=E_+^{(0)}-E_-^{(0)}=\sqrt{\eta^2+8\xi_\mathrm{SO}^2},\label{CF}
\end{eqnarray}
with $\eta=\xi_\mathrm{SO}+(M-L)\lambda^2$, and $F(\beta)$ is the correction when $V=\mathcal{E}_zz$ [see Appendix \ref{S2} for details].

\begin{figure}
  \includegraphics[width=\linewidth]{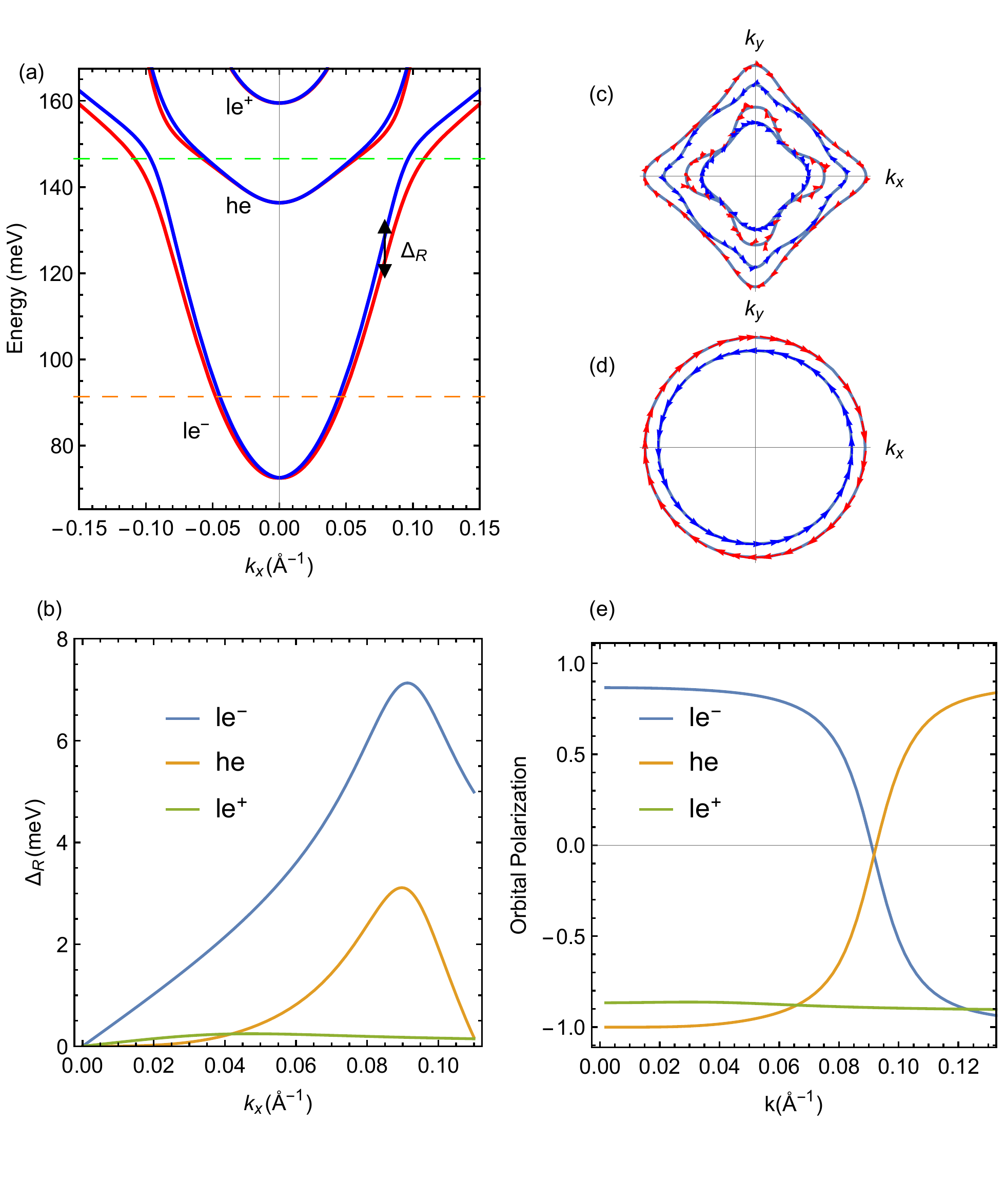}
  \caption{(a) Bandstructure obtained from the full effective Hamiltonian includes two light electron bands ($le^\pm$) sandwiched by a heavy electron band ($he$).  (b) The corresponding spin-splitting as function of momentum. (c)-(d) Energy contours and spin textures at different Fermi level: near bottom edge of the $le^-$ band (orange dashed line), the spin splitting is of linear Rashba type with characteristic spin orientations along the isotropic energy contours (d); whereas close to the crossing region (green dashed line), substantial cubic and higher order spin splitting yields anisotropic contours and deformed spin textures(c).  (e) Orbital polarization defined in Eq. (\ref{orbpola}) represents the orbital character in each band: $d_{xy}$ and $d_{yz/zx}$ are predominant around $\Gamma$-point in the $le^-$ and $he$ bands, respectively, whereas they are more equalized close to the crossing point.}
  \label{Fig3}
\end{figure}

Rearranging the sequence of the six eigenvecctors as $\{\psi_+^{\urow},\psi_+^{\drow},\psi_0^{\urow},\psi_0^{\drow},\psi_-^{\urow}, \psi_-^{\drow}\}$,
 and by projecting the full Hamiltonian to the spin-orbital space of these vectors, we obtain an effective Hamiltonian beyond the $\Gamma$-point in block form as (see Appendix \ref{S3} for the full expression)
\begin{eqnarray}\label{Heff}
\mathcal{H}=
\begin{bmatrix}
\mathcal{H}_{+}&\mathcal{G}_{3}&\mathcal{G}_{1}\\
\mathcal{G}_{3}^\dag &\mathcal{H}_{0}&\mathcal{G}_{2}\\
\mathcal{G}_{1}^\dag &\mathcal{G}_{2}^\dag &\mathcal{H}_{-}
\end{bmatrix}.
\end{eqnarray}
In the above, the diagonal blocks are given by (for $\tau=0,\pm$)
\begin{eqnarray}\label{hpm1}
\mathcal{H}_{\tau}=E_\tau+\frac{\hbar^2 {\bm k}^2}{2m^*_\tau}+\alpha_\tau ({\bm k}\times {\bm z})\cdot {\bm\sigma},
\end{eqnarray}
which are characterised by bottom band edges $E_\tau$, effective masses $m^*_\tau$, and Rashba parameters $\alpha_\tau$. Here, ${\bm k}=(k_x,k_y)$ is the electron momentum in the $x-y$ plane, $\bm \sigma$ is the vector of the Pauli matrices, and $\bm z$ is the unit vector perpendicular to the STO slab.

First, we look at the effective masses given by 
$
\frac{1}{m^*_{0}}=\frac{(L+M)}{\hbar^2}, \frac{1}{m^*_\pm}= \frac{1}{m^*_0}+\frac{(M-L)}{2\hbar^2}\left(1\pm\frac{\eta}{\Delta_I}\right). $
Recall that $M>L$, and $\eta/\Delta_I<1$, we have $m^*_\pm< m^*_0$,  which indicates that $\mathcal{H}_\pm$ describe light-electron (denoted as $le^\pm$) bands, whereas $\mathcal{H}_0$ corresponds to the  heavy-electron ($he$)  band. Furthermore, it can be verified that $E_+>E_0>E_-$, so that the two $le$-bands are sandwiched by the $he$-band. This arrangement leads to the anti-crossing between the bottom ($le^-$)and the middle ($he$) bands [see Fig. \ref{Fig3}(a)].

\subsection{Linear Rashba SOC } Now let us consider the expressions of linear RSOC given by
\begin{eqnarray}\label{RSOC}
\alpha_0=0,\ \ \alpha_{\pm}&&=\pm \frac{4\xi_\mathrm{SO}{\bar \gamma}_\pm}{\Delta_I},
\end{eqnarray}
where $\Delta_I$ is given in Eq. (\ref{CF}) regarded as effective crystal field, and
\begin{eqnarray}
{\bar \gamma}_\pm&&=\int{dz \gamma(z)|f_\pm(z)|^2},\label{ISBeff}
\end{eqnarray}
is considered as the effective inversion-symmetry breaking, with the integration taken over the STO slab. In the above, $|f(z)|^2$ is the spatial distribution function, thus ${\bar \gamma}$ is the average ISB parameter across the STO slab. As the ISB constitutes an interface effect, it is only finite near the interface layer, and becomes vanishingly small in the bulk \cite{RN402,RN403}. Thus, Eq. (\ref{ISBeff}) indicates that the localization of electron at the interface is significant for inducing large effective ISB, which can be achieved by having a large confinement effect induced by external electric field and/or charge density \cite{STOdensity}. To obtain a simple analytical  expression of the average ISB parameter, we consider the case where the ISB is finite only at the interface layer and zero elsewhere in the bulk, i.e., $\gamma(z)=\gamma_0\theta(a-z)$, where $\theta(x)$ is the Heaviside step function, and $a=3.870 \mathrm{\AA}$ is the lattice constant of STO crystal. In this case, the average ISB parameter evaluates as
\begin{eqnarray}\label{ISB1}
{\bar \gamma}_\pm/\gamma_0=\frac{ a^3}{M_\pm}  \mathcal{E}_z+2\pi^2\left(\frac{4}{3 M_\pm^2}\right)^{2/3}\frac{a^3}{d^2} \mathcal{E}_z^{1/3},
\end{eqnarray}
in which $M_\pm=M/2\left(1\pm\frac{\eta}{\Delta_I}\right)$. Eq. (\ref{ISB1}) shows that the average ISB, and thus the RSOC, is scaled with the interface confining electric field $\mathcal{E}_z$ to the first order as depicted in Fig. \ref{Fig4}(c). 

Furthermore, it can be seen that reducing the QW width $d$ (STO thickness) will increase the effective ISB due to the wavefunction localization near the interface. However,  the effective crystal field in the thin film limit reads as 
\begin{eqnarray}
\Delta_I\approx \frac{(M-L)\pi^2}{d^2},
\end{eqnarray}
 which also increases with decreasing $d$. This trend is ascribed to the quantum confinement effect in the STO quantum well, which causes the subbands further split in the thin STO limit.  From Eq. (\ref{RSOC}), these competing trends above between the effective ISB and effective crystal field lead to non-monotonic thickness dependence of the RSOC [see Fig. \ref{Fig4}(a)]. In general, the RSOC diminishes with reducing STO thickness, which is one of our main results.

 \noindent On the other hand, at large STO thickness, the effective crystal field reduces to $\Delta_I\approx 3\xi_\mathrm{SO}$, and ${\bar \gamma}_\pm\approx \frac{a^3\mathcal{E}_z}{M_\pm}$, which result in a saturation of the linear Rashba to $\alpha_{\pm}\approx\frac{4\gamma_0 a^3\mathcal{E}_z}{3M_\pm}$ as shown in Fig. \ref{Fig4}(a).  It is interesting to see that in the large thickness limit, the saturation value of RSOC has a weak dependence on the atomic SOC $\xi_\mathrm{SO}$. However, this does not deny the vital role of the atomic SOC, as the RSOC expression in Eq. (\ref{RSOC}) would vanish if  $\xi_\mathrm{SO}$ is set to zero in the first place. Assuming typical parameter values, the linear Rashba coefficient is found to be in the range of $1-10$ meV\AA,  corresponding to spin-splitting of $1-10$ meV,  which is consistent with previous computational \cite{RN339,PhysRevB.88.041302,Zhou:PRB15} and experimental works \cite{RN337,RN406,RN407,RN372}.

We note that in other transition metal oxides such as $\mathrm{SrIrO_3}$ \cite{SrIrO3:review}, the atomic SOC is much stronger than that in the $\mathrm{SrTiO_3}$. However, the crystal field in such oxides is also much larger at the same time \cite{SrIrO3:review}. Therefore, following Eq. (\ref{RSOC}), the strong atomic SOC oxides  does not always translate into a strong Rashba SOC. Experimentally, it has been shown that the Rashba SOC in $\mathrm{SrIrO_3}$ is about 5 meV\AA \cite{SrIrO3:RSOC}, a strength which is comparable to that in the $\mathrm{SrTiO_3}$.


\subsection{ Cubic Rashba SOC }

\begin{figure}
  \includegraphics[width=\linewidth]{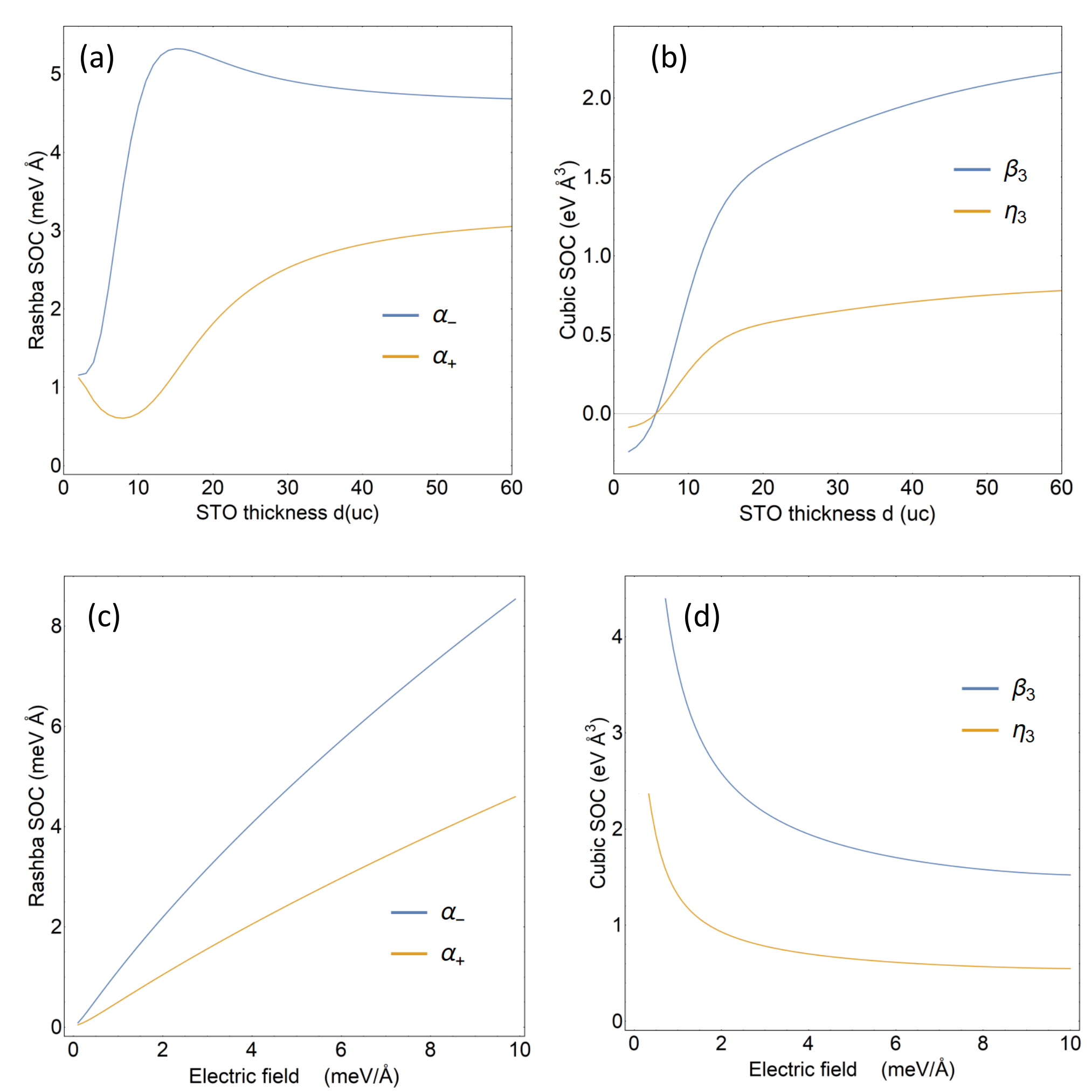}
  \caption{Dependence of the linear and cubic SOC parameters on STO thickness $d$ when $\mathcal{E}_z=5$ meV/\AA (a)-(b), and  on electric field $\mathcal{E}_z$ when $d=50$ \AA (c)-(d). $\alpha_\pm$ are the linear RSOC parameters of the light electron bands ($le^\pm$), respectively, and $\beta_3,\eta_3$ are the cubic SOC parameters of the heavy electron band ($he$).}
  \label{Fig4}
\end{figure}

 Now we turn to discuss role of the off-diagonal elements $\mathcal{G}$ in Eq. (\ref{Heff}), which describe the coupling between the $m_j=\pm 3/2$ and $m_j=\pm 1/2$ states. It has been shown that the above coupling is the prerequisite for the presence of cubic SOC in the $m_j=3/2$ states \cite{winkler2003spin}, which in our case is the $he$ bands. Indeed, by reducing the $6\times 6$ Hamiltonian (\ref{Heff}) to three $2\times 2$ matrices by means of quasi-degenerate partitioning \cite{winkler2003spin}, we obtain cubic corrections as
\begin{eqnarray}
\delta \mathcal{H}^\mathrm{cubic}_{le^-}=&&\beta_3 (k_x^2-k_y^2)(k_y \sigma_x-k_x \sigma_y)\nonumber\\
&&-\eta_3 k_xk_y(k_x\sigma_x-k_y\sigma_y),\\
\delta \mathcal{H}^\mathrm{cubic}_{he}=&&\beta_3 (k_x^2-k_y^2)(k_y \sigma_x-k_x \sigma_y)\nonumber\\
&&+\eta_3 k_xk_y(k_x\sigma_x-k_y\sigma_y),
\end{eqnarray}
 in which the $k$-cubed SOC parameters are given by
\begin{eqnarray}\label{cubic}
\beta_3=\frac{(M-L)\xi_\mathrm{SO}R}{\Delta_I\Delta_-},\ \ \eta_3=-\frac{2N\xi_\mathrm{SO}R}{\Delta_I\Delta_-},
\end{eqnarray}
in which $R=(2\gamma_{-0}-N\kappa_{-0})\rho_{-0}, \Delta_-={E_0-E_-}$, where we defined $\gamma_{ij}=\int{\gamma(z)f_if_j}, \kappa_{ij}=\int{f_i\partial_z f_j}$, and $\rho_{ij}=\int{f_if_j}$ [see Appendix \ref{S4}]. In Eq. (\ref{cubic}), there are two cubic RSOC terms with different anisotropy in $k$-space. The first term depends on the mass difference, i.e., $M-L$, which is responsible for the mass anisotropy in the bulk STO. The second term stems from the the bulk inter-orbital hopping $N$, which is responsible for the asymmetry in the bulk. The cubic SOC parameters have a similar thickness dependence as in the linear case, in that they saturate at large thickness limit as shown in Fig. \ref{Fig4} (b). However, the cubic SOC diminishes with increasing electric field, in contrast to the opposite trend shown by the linear RSOC plotted in Figs. \ref{Fig4}(c) and (d). With typical parameter values,, the cubic RSOC is estimated to be in the range $1-4$ eV$\AA^3$, which is in agreement with previous experimental and computional works \cite{RN339,RN408}. Furthermore, the strength of cubic SOC has been shown to decrease at high electron density \cite{RN408}, a trend which is consistent with our theory taking into consideration the linear correlation between  the interface electric field and electron density \cite{STOdensity}.

\subsection{Orbital polarization} In the above, we obtain three pairs of bands, each of which is a mix of $d_{xy}, d_{yz}$, and $d_{zx}$ orbitals. To analyze the contribution of these orbitals in each band, we introduce the orbital occupancy which is obtained by projecting the eigenstates of the effective Hamiltonian \ref{Heff} onto the $t_{2g}$ basis states (see Appendix \ref{S5} for details). Denote $\mathcal{N}_i^\tau$ as the occupancy of orbital $i=xy,yz,zx$ in band $\tau=le^{\pm},he$, we can define the orbital polarization as
\begin{eqnarray}\label{orbpola}
\mathcal{P}_\tau=\frac{\mathcal{N}_{xy}^{\tau}-\mathcal{N}_{yz+zx}^{\tau}}{\mathcal{N}_{xy}^{\tau}+\mathcal{N}_{yz+zx}^{\tau}},
\end{eqnarray}
which characterizes the contribution of the preferable $d_{xy}$ in the interfacial 2DEG channel. In Fig. \ref{Fig3}(e), the orbital polarization is depicted, which shows drastically different contribution of the orbitals in each band at different energy region.  Around the $\Gamma$-point, the orbitals are well polarized with $d_{xy}$ and $d_{yz/zx}$ predominant in the two lowest bands, respectively. In this case, the orbital polarization can be obtained as
$\mathcal{P}_{he}=-1, \mathcal{P}_{le^\pm}=\mp\frac{\eta}{\Delta_I}.$
Close to the crossing region, stronger mixing between the $d_{xy}$-like and $d_{yz/zx}$-like bands will neutralize the orbital polarization. This trend is consistent with previous computational works \cite{orbital1,orbital2}, where it has been shown that the $d_{xy}$ orbital filling is less at higher energy region. At the same time,
 the relative contribution of the different orbitals can be controlled by tuning the heterostructure's parameters such as STO thickness. For example, the $d_{xy}$ orbital contribution becomes more dominant in the lowest $le^-$ band  when the STO thickness is reduced, and approaches $\mathcal{P}_{le^-}\approx 1$ in the thin limit, whereas its weight reduces as 
\begin{eqnarray}
\mathcal{P}_{le^-}\approx \frac{1}{3}\left[1+\frac{8(M-L)\pi^2}{9\xi_\mathrm{SO}d^2}\right],
\end{eqnarray}
 in the thick limit. 

The thickness dependence of the $d_{xy}$ occupancy is in line with the diminishing of the RSOC in the thin limit. The RSOC originates from the ISB, which represents the hopping between the $d_{xy}$ and $d_{yz/zx}$ at the interface as shown in Eq. (\ref{ISB}). This means that only bands that occupied by both $d_{xy}$ and $d_{yz/zx}$ orbitals will show the linear RSOC. Therefore, in thin STO slabs, the $d_{yz/zx}$ disappears in the $le^-$ bands resulting in the vanishing of the hopping as well as the RSOC. Similarly, since there is no $d_{xy}$ in the $he$ bands, the linear RSOC is absent in these bands at all STO thickness. Instead, the $he$ band shows cubic SOC as discussed in the previous section.

\section{ Conclusion}
In this work, we formulate the analytical Hamiltonian of Rashba splitting in the $t_{2g}$ bands of STO heterostructure based on ${\bf k\cdot p}$ formalism.  First, we express the Rashba parameter as functions of the heterostructure parameters and show that it is thickness dependent. Strikingly, the RSOC diminishes as the STO thickness reduces, and it saturates in thick STO layer. The hybridization between the orbitals results in orbital-dependent Rashba spin splittings. Explicitly, the linear Rashba SOC is associated with the $d_{xy}$ orbital which preferably occupies the lowest band at low energy region. On the other hand, at higher energy crossing region, the cubic Rashba SOC accompanies the enriched $d_{yz/zx}$ orbitals. These results reveal the capability to control the spin-orbit coupling and orbital selection in $\mathrm{LaAlO_3/SrTiO_3}$ heterostructures for spintronics applications and devices.

\begin{acknowledgments}
This work is supported by the Ministry of Educations (MOE2015-T2-2-065, MOE2015-T2-1-099, MOE2015-T2-2-147, and MOE2017-T2-1-135), Singapore National Research Foundation under its Competitive Research Funding (NRF-CRP 8-2011-06 and NRF-CRP15-2015-01) and under its Medium Sized Centre Programme (Centre for Advanced 2D Materials and Graphene Research Centre), and FRC (R-144-000-379-114 and R-144-000-368-112).
\end{acknowledgments}


\begin{thebibliography}{48}%
\makeatletter
\providecommand \@ifxundefined [1]{%
 \@ifx{#1\undefined}
}%
\providecommand \@ifnum [1]{%
 \ifnum #1\expandafter \@firstoftwo
 \else \expandafter \@secondoftwo
 \fi
}%
\providecommand \@ifx [1]{%
 \ifx #1\expandafter \@firstoftwo
 \else \expandafter \@secondoftwo
 \fi
}%
\providecommand \natexlab [1]{#1}%
\providecommand \enquote  [1]{``#1''}%
\providecommand \bibnamefont  [1]{#1}%
\providecommand \bibfnamefont [1]{#1}%
\providecommand \citenamefont [1]{#1}%
\providecommand \href@noop [0]{\@secondoftwo}%
\providecommand \href [0]{\begingroup \@sanitize@url \@href}%
\providecommand \@href[1]{\@@startlink{#1}\@@href}%
\providecommand \@@href[1]{\endgroup#1\@@endlink}%
\providecommand \@sanitize@url [0]{\catcode `\\12\catcode `\$12\catcode
  `\&12\catcode `\#12\catcode `\^12\catcode `\_12\catcode `\%12\relax}%
\providecommand \@@startlink[1]{}%
\providecommand \@@endlink[0]{}%
\providecommand \url  [0]{\begingroup\@sanitize@url \@url }%
\providecommand \@url [1]{\endgroup\@href {#1}{\urlprefix }}%
\providecommand \urlprefix  [0]{URL }%
\providecommand \Eprint [0]{\href }%
\providecommand \doibase [0]{http://dx.doi.org/}%
\providecommand \selectlanguage [0]{\@gobble}%
\providecommand \bibinfo  [0]{\@secondoftwo}%
\providecommand \bibfield  [0]{\@secondoftwo}%
\providecommand \translation [1]{[#1]}%
\providecommand \BibitemOpen [0]{}%
\providecommand \bibitemStop [0]{}%
\providecommand \bibitemNoStop [0]{.\EOS\space}%
\providecommand \EOS [0]{\spacefactor3000\relax}%
\providecommand \BibitemShut  [1]{\csname bibitem#1\endcsname}%
\let\auto@bib@innerbib\@empty
\bibitem [{\citenamefont {Huang}\ \emph
  {et~al.}(2018{\natexlab{a}})\citenamefont {Huang}, \citenamefont {Ariando},
  \citenamefont {Renshaw~Wang}, \citenamefont {Rusydi}, \citenamefont {Chen},
  \citenamefont {Yang},\ and\ \citenamefont {Venkatesan}}]{Andrivo:2018}%
  \BibitemOpen
  \bibfield  {author} {\bibinfo {author} {\bibfnamefont {Z.}~\bibnamefont
  {Huang}}, \bibinfo {author} {\bibnamefont {Ariando}}, \bibinfo {author}
  {\bibfnamefont {X.}~\bibnamefont {Renshaw~Wang}}, \bibinfo {author}
  {\bibfnamefont {A.}~\bibnamefont {Rusydi}}, \bibinfo {author} {\bibfnamefont
  {J.}~\bibnamefont {Chen}}, \bibinfo {author} {\bibfnamefont {H.}~\bibnamefont
  {Yang}}, \ and\ \bibinfo {author} {\bibfnamefont {T.}~\bibnamefont
  {Venkatesan}},\ }\href {\doibase 10.1002/adma.201802439} {\bibfield
  {journal} {\bibinfo  {journal} {Adv. Mater.}\ }\textbf {\bibinfo {volume}
  {0}},\ \bibinfo {pages} {1802439} (\bibinfo {year}
  {2018}{\natexlab{a}})}\BibitemShut {NoStop}%
\bibitem [{\citenamefont {Ohtomo}\ and\ \citenamefont
  {Hwang}(2004{\natexlab{a}})}]{ohtomo2004high}%
  \BibitemOpen
  \bibfield  {author} {\bibinfo {author} {\bibfnamefont {A.}~\bibnamefont
  {Ohtomo}}\ and\ \bibinfo {author} {\bibfnamefont {H.}~\bibnamefont {Hwang}},\
  }\href {https://doi.org/10.1038/nature02308} {\bibfield  {journal} {\bibinfo
  {journal} {Nature}\ }\textbf {\bibinfo {volume} {427}},\ \bibinfo {pages}
  {423} (\bibinfo {year} {2004}{\natexlab{a}})}\BibitemShut {NoStop}%
\bibitem [{\citenamefont {Annadi}\ \emph {et~al.}(2013)\citenamefont {Annadi},
  \citenamefont {Zhang}, \citenamefont {Wang}, \citenamefont {Tuzla},
  \citenamefont {Gopinadhan}, \citenamefont {L{\"u}}, \citenamefont {Barman},
  \citenamefont {Liu}, \citenamefont {Srivastava}, \citenamefont {Saha} \emph
  {et~al.}}]{2DEG1}%
  \BibitemOpen
  \bibfield  {author} {\bibinfo {author} {\bibfnamefont {A.}~\bibnamefont
  {Annadi}}, \bibinfo {author} {\bibfnamefont {Q.}~\bibnamefont {Zhang}},
  \bibinfo {author} {\bibfnamefont {X.~R.}\ \bibnamefont {Wang}}, \bibinfo
  {author} {\bibfnamefont {N.}~\bibnamefont {Tuzla}}, \bibinfo {author}
  {\bibfnamefont {K.}~\bibnamefont {Gopinadhan}}, \bibinfo {author}
  {\bibfnamefont {W.}~\bibnamefont {L{\"u}}}, \bibinfo {author} {\bibfnamefont
  {A.~R.}\ \bibnamefont {Barman}}, \bibinfo {author} {\bibfnamefont
  {Z.}~\bibnamefont {Liu}}, \bibinfo {author} {\bibfnamefont {A.}~\bibnamefont
  {Srivastava}}, \bibinfo {author} {\bibfnamefont {S.}~\bibnamefont {Saha}},
  \emph {et~al.},\ }\href {https://doi.org/10.1038/ncomms2804} {\bibfield
  {journal} {\bibinfo  {journal} {Nat. Commun.}\ }\textbf {\bibinfo {volume}
  {4}},\ \bibinfo {pages} {1838} (\bibinfo {year} {2013})}\BibitemShut
  {NoStop}%
\bibitem [{\citenamefont {Siemons}\ \emph {et~al.}(2007)\citenamefont
  {Siemons}, \citenamefont {Koster}, \citenamefont {Yamamoto}, \citenamefont
  {Harrison}, \citenamefont {Lucovsky}, \citenamefont {Geballe}, \citenamefont
  {Blank},\ and\ \citenamefont {Beasley}}]{2DEG2}%
  \BibitemOpen
  \bibfield  {author} {\bibinfo {author} {\bibfnamefont {W.}~\bibnamefont
  {Siemons}}, \bibinfo {author} {\bibfnamefont {G.}~\bibnamefont {Koster}},
  \bibinfo {author} {\bibfnamefont {H.}~\bibnamefont {Yamamoto}}, \bibinfo
  {author} {\bibfnamefont {W.~A.}\ \bibnamefont {Harrison}}, \bibinfo {author}
  {\bibfnamefont {G.}~\bibnamefont {Lucovsky}}, \bibinfo {author}
  {\bibfnamefont {T.~H.}\ \bibnamefont {Geballe}}, \bibinfo {author}
  {\bibfnamefont {D.~H.}\ \bibnamefont {Blank}}, \ and\ \bibinfo {author}
  {\bibfnamefont {M.~R.}\ \bibnamefont {Beasley}},\ }\href
  {https://doi.org/10.1103/PhysRevLett.98.196802} {\bibfield  {journal}
  {\bibinfo  {journal} {Phys. Rev. Lett.}\ }\textbf {\bibinfo {volume} {98}},\
  \bibinfo {pages} {196802} (\bibinfo {year} {2007})}\BibitemShut {NoStop}%
\bibitem [{\citenamefont {Popovi{\'c}}\ \emph {et~al.}(2008)\citenamefont
  {Popovi{\'c}}, \citenamefont {Satpathy},\ and\ \citenamefont
  {Martin}}]{2DEG3}%
  \BibitemOpen
  \bibfield  {author} {\bibinfo {author} {\bibfnamefont {Z.~S.}\ \bibnamefont
  {Popovi{\'c}}}, \bibinfo {author} {\bibfnamefont {S.}~\bibnamefont
  {Satpathy}}, \ and\ \bibinfo {author} {\bibfnamefont {R.~M.}\ \bibnamefont
  {Martin}},\ }\href {https://doi.org/10.1103/PhysRevLett.101.256801}
  {\bibfield  {journal} {\bibinfo  {journal} {Phys. Rev. Lett.}\ }\textbf
  {\bibinfo {volume} {101}},\ \bibinfo {pages} {256801} (\bibinfo {year}
  {2008})}\BibitemShut {NoStop}%
\bibitem [{\citenamefont {Brinkman}\ \emph {et~al.}(2007)\citenamefont
  {Brinkman}, \citenamefont {Huijben}, \citenamefont {Van~Zalk}, \citenamefont
  {Huijben}, \citenamefont {Zeitler}, \citenamefont {Maan}, \citenamefont
  {van~der Wiel}, \citenamefont {Rijnders}, \citenamefont {Blank},\ and\
  \citenamefont {Hilgenkamp}}]{brinkman2007magnetic}%
  \BibitemOpen
  \bibfield  {author} {\bibinfo {author} {\bibfnamefont {A.}~\bibnamefont
  {Brinkman}}, \bibinfo {author} {\bibfnamefont {M.}~\bibnamefont {Huijben}},
  \bibinfo {author} {\bibfnamefont {M.}~\bibnamefont {Van~Zalk}}, \bibinfo
  {author} {\bibfnamefont {J.}~\bibnamefont {Huijben}}, \bibinfo {author}
  {\bibfnamefont {U.}~\bibnamefont {Zeitler}}, \bibinfo {author} {\bibfnamefont
  {J.}~\bibnamefont {Maan}}, \bibinfo {author} {\bibfnamefont {W.~G.}\
  \bibnamefont {van~der Wiel}}, \bibinfo {author} {\bibfnamefont
  {G.}~\bibnamefont {Rijnders}}, \bibinfo {author} {\bibfnamefont {D.~H.}\
  \bibnamefont {Blank}}, \ and\ \bibinfo {author} {\bibfnamefont
  {H.}~\bibnamefont {Hilgenkamp}},\ }\href {https://doi.org/10.1038/nmat1931}
  {\bibfield  {journal} {\bibinfo  {journal} {Nat. Mater.}\ }\textbf {\bibinfo
  {volume} {6}},\ \bibinfo {pages} {493} (\bibinfo {year} {2007})}\BibitemShut
  {NoStop}%
\bibitem [{\citenamefont {Kalisky}\ \emph {et~al.}(2012)\citenamefont
  {Kalisky}, \citenamefont {Bert}, \citenamefont {Klopfer}, \citenamefont
  {Bell}, \citenamefont {Sato}, \citenamefont {Hosoda}, \citenamefont {Hikita},
  \citenamefont {Hwang},\ and\ \citenamefont {Moler}}]{kalisky2012critical}%
  \BibitemOpen
  \bibfield  {author} {\bibinfo {author} {\bibfnamefont {B.}~\bibnamefont
  {Kalisky}}, \bibinfo {author} {\bibfnamefont {J.~A.}\ \bibnamefont {Bert}},
  \bibinfo {author} {\bibfnamefont {B.~B.}\ \bibnamefont {Klopfer}}, \bibinfo
  {author} {\bibfnamefont {C.}~\bibnamefont {Bell}}, \bibinfo {author}
  {\bibfnamefont {H.~K.}\ \bibnamefont {Sato}}, \bibinfo {author}
  {\bibfnamefont {M.}~\bibnamefont {Hosoda}}, \bibinfo {author} {\bibfnamefont
  {Y.}~\bibnamefont {Hikita}}, \bibinfo {author} {\bibfnamefont {H.~Y.}\
  \bibnamefont {Hwang}}, \ and\ \bibinfo {author} {\bibfnamefont {K.~A.}\
  \bibnamefont {Moler}},\ }\href {https://doi.org/10.1038/ncomms1931}
  {\bibfield  {journal} {\bibinfo  {journal} {Nat. Commun.}\ }\textbf {\bibinfo
  {volume} {3}},\ \bibinfo {pages} {922} (\bibinfo {year} {2012})}\BibitemShut
  {NoStop}%
\bibitem [{\citenamefont {Li}\ \emph {et~al.}(2011)\citenamefont {Li},
  \citenamefont {Richter}, \citenamefont {Mannhart},\ and\ \citenamefont
  {Ashoori}}]{li2011coexistence}%
  \BibitemOpen
  \bibfield  {author} {\bibinfo {author} {\bibfnamefont {L.}~\bibnamefont
  {Li}}, \bibinfo {author} {\bibfnamefont {C.}~\bibnamefont {Richter}},
  \bibinfo {author} {\bibfnamefont {J.}~\bibnamefont {Mannhart}}, \ and\
  \bibinfo {author} {\bibfnamefont {R.}~\bibnamefont {Ashoori}},\ }\href
  {https://doi.org/10.1038/nphys2080} {\bibfield  {journal} {\bibinfo
  {journal} {Nat. Phys.}\ }\textbf {\bibinfo {volume} {7}},\ \bibinfo {pages}
  {762} (\bibinfo {year} {2011})}\BibitemShut {NoStop}%
\bibitem [{\citenamefont {Dikin}\ \emph {et~al.}(2011)\citenamefont {Dikin},
  \citenamefont {Mehta}, \citenamefont {Bark}, \citenamefont {Folkman},
  \citenamefont {Eom},\ and\ \citenamefont
  {Chandrasekhar}}]{dikin2011coexistence}%
  \BibitemOpen
  \bibfield  {author} {\bibinfo {author} {\bibfnamefont {D.}~\bibnamefont
  {Dikin}}, \bibinfo {author} {\bibfnamefont {M.}~\bibnamefont {Mehta}},
  \bibinfo {author} {\bibfnamefont {C.}~\bibnamefont {Bark}}, \bibinfo {author}
  {\bibfnamefont {C.}~\bibnamefont {Folkman}}, \bibinfo {author} {\bibfnamefont
  {C.}~\bibnamefont {Eom}}, \ and\ \bibinfo {author} {\bibfnamefont
  {V.}~\bibnamefont {Chandrasekhar}},\ }\href
  {https://doi.org/10.1103/PhysRevLett.107.056802} {\bibfield  {journal}
  {\bibinfo  {journal} {Phys. Rev. Lett.}\ }\textbf {\bibinfo {volume} {107}},\
  \bibinfo {pages} {056802} (\bibinfo {year} {2011})}\BibitemShut {NoStop}%
\bibitem [{\citenamefont {Yang}\ \emph {et~al.}(2018)\citenamefont {Yang},
  \citenamefont {Ariando}, \citenamefont {Zhou}, \citenamefont {Asmara},
  \citenamefont {Krüger}, \citenamefont {Yu}, \citenamefont {Wang},
  \citenamefont {Sanchez-Hanke}, \citenamefont {Feng}, \citenamefont
  {Venkatesan},\ and\ \citenamefont {Rusydi}}]{Yangming:2018}%
  \BibitemOpen
  \bibfield  {author} {\bibinfo {author} {\bibfnamefont {M.}~\bibnamefont
  {Yang}}, \bibinfo {author} {\bibnamefont {Ariando}}, \bibinfo {author}
  {\bibfnamefont {J.}~\bibnamefont {Zhou}}, \bibinfo {author} {\bibfnamefont
  {T.~C.}\ \bibnamefont {Asmara}}, \bibinfo {author} {\bibfnamefont
  {P.}~\bibnamefont {Krüger}}, \bibinfo {author} {\bibfnamefont {X.~J.}\
  \bibnamefont {Yu}}, \bibinfo {author} {\bibfnamefont {X.}~\bibnamefont
  {Wang}}, \bibinfo {author} {\bibfnamefont {C.}~\bibnamefont {Sanchez-Hanke}},
  \bibinfo {author} {\bibfnamefont {Y.~P.}\ \bibnamefont {Feng}}, \bibinfo
  {author} {\bibfnamefont {T.}~\bibnamefont {Venkatesan}}, \ and\ \bibinfo
  {author} {\bibfnamefont {A.}~\bibnamefont {Rusydi}},\ }\href
  {https://doi.org/10.1021/acsami.7b12945} {\bibfield  {journal} {\bibinfo
  {journal} {ACS Appl. Mater. Interfaces}\ }\textbf {\bibinfo {volume} {10}},\
  \bibinfo {pages} {9774} (\bibinfo {year} {2018})}\BibitemShut {NoStop}%
\bibitem [{\citenamefont {Hwang}\ \emph {et~al.}(2012)\citenamefont {Hwang},
  \citenamefont {Iwasa}, \citenamefont {Kawasaki}, \citenamefont {Keimer},
  \citenamefont {Nagaosa},\ and\ \citenamefont {Tokura}}]{hwang2012emergent}%
  \BibitemOpen
  \bibfield  {author} {\bibinfo {author} {\bibfnamefont {H.~Y.}\ \bibnamefont
  {Hwang}}, \bibinfo {author} {\bibfnamefont {Y.}~\bibnamefont {Iwasa}},
  \bibinfo {author} {\bibfnamefont {M.}~\bibnamefont {Kawasaki}}, \bibinfo
  {author} {\bibfnamefont {B.}~\bibnamefont {Keimer}}, \bibinfo {author}
  {\bibfnamefont {N.}~\bibnamefont {Nagaosa}}, \ and\ \bibinfo {author}
  {\bibfnamefont {Y.}~\bibnamefont {Tokura}},\ }\href
  {https://doi.org/10.1038/nmat3223} {\bibfield  {journal} {\bibinfo  {journal}
  {Nat. Mater.}\ }\textbf {\bibinfo {volume} {11}},\ \bibinfo {pages} {103}
  (\bibinfo {year} {2012})}\BibitemShut {NoStop}%
\bibitem [{\citenamefont {Shalom}\ \emph {et~al.}(2010)\citenamefont {Shalom},
  \citenamefont {Sachs}, \citenamefont {Rakhmilevitch}, \citenamefont
  {Palevski},\ and\ \citenamefont {Dagan}}]{RN338}%
  \BibitemOpen
  \bibfield  {author} {\bibinfo {author} {\bibfnamefont {M.~B.}\ \bibnamefont
  {Shalom}}, \bibinfo {author} {\bibfnamefont {M.}~\bibnamefont {Sachs}},
  \bibinfo {author} {\bibfnamefont {D.}~\bibnamefont {Rakhmilevitch}}, \bibinfo
  {author} {\bibfnamefont {A.}~\bibnamefont {Palevski}}, \ and\ \bibinfo
  {author} {\bibfnamefont {Y.}~\bibnamefont {Dagan}},\ }\href
  {https://doi.org/10.1103/PhysRevLett.104.126802} {\bibfield  {journal}
  {\bibinfo  {journal} {Phys. Rev. Lett.}\ }\textbf {\bibinfo {volume} {104}},\
  \bibinfo {pages} {126802} (\bibinfo {year} {2010})}\BibitemShut {NoStop}%
\bibitem [{\citenamefont {Zhong}\ \emph {et~al.}(2013)\citenamefont {Zhong},
  \citenamefont {T\'{o}th},\ and\ \citenamefont {Held}}]{RN339}%
  \BibitemOpen
  \bibfield  {author} {\bibinfo {author} {\bibfnamefont {Z.}~\bibnamefont
  {Zhong}}, \bibinfo {author} {\bibfnamefont {A.}~\bibnamefont {T\'{o}th}}, \
  and\ \bibinfo {author} {\bibfnamefont {K.}~\bibnamefont {Held}},\ }\href
  {https://doi.org/10.1103/PhysRevB.87.161102} {\bibfield  {journal} {\bibinfo
  {journal} {Phys. Rev. B}\ }\textbf {\bibinfo {volume} {87}},\ \bibinfo
  {pages} {161102} (\bibinfo {year} {2013})}\BibitemShut {NoStop}%
\bibitem [{\citenamefont {Caviglia}\ \emph {et~al.}(2010)\citenamefont
  {Caviglia}, \citenamefont {Gabay}, \citenamefont {Gariglio}, \citenamefont
  {Reyren}, \citenamefont {Cancellieri},\ and\ \citenamefont
  {Triscone}}]{RN337}%
  \BibitemOpen
  \bibfield  {author} {\bibinfo {author} {\bibfnamefont {A.}~\bibnamefont
  {Caviglia}}, \bibinfo {author} {\bibfnamefont {M.}~\bibnamefont {Gabay}},
  \bibinfo {author} {\bibfnamefont {S.}~\bibnamefont {Gariglio}}, \bibinfo
  {author} {\bibfnamefont {N.}~\bibnamefont {Reyren}}, \bibinfo {author}
  {\bibfnamefont {C.}~\bibnamefont {Cancellieri}}, \ and\ \bibinfo {author}
  {\bibfnamefont {J.-M.}\ \bibnamefont {Triscone}},\ }\href@noop {} {\bibfield
  {journal} {\bibinfo  {journal} {Phys. Rev. Lett.}\ }\textbf {\bibinfo
  {volume} {104}},\ \bibinfo {pages} {126803} (\bibinfo {year}
  {2010})}\BibitemShut {NoStop}%
\bibitem [{\citenamefont {Fête}\ \emph {et~al.}(2012)\citenamefont {Fête},
  \citenamefont {Gariglio}, \citenamefont {Caviglia}, \citenamefont
  {Triscone},\ and\ \citenamefont {Gabay}}]{RN406}%
  \BibitemOpen
  \bibfield  {author} {\bibinfo {author} {\bibfnamefont {A.}~\bibnamefont
  {Fête}}, \bibinfo {author} {\bibfnamefont {S.}~\bibnamefont {Gariglio}},
  \bibinfo {author} {\bibfnamefont {A.~D.}\ \bibnamefont {Caviglia}}, \bibinfo
  {author} {\bibfnamefont {J.~M.}\ \bibnamefont {Triscone}}, \ and\ \bibinfo
  {author} {\bibfnamefont {M.}~\bibnamefont {Gabay}},\ }\href
  {https://link.aps.org/doi/10.1103/PhysRevB.86.201105} {\bibfield  {journal}
  {\bibinfo  {journal} {Phys. Rev. B}\ }\textbf {\bibinfo {volume} {86}},\
  \bibinfo {pages} {201105} (\bibinfo {year} {2012})}\BibitemShut {NoStop}%
\bibitem [{\citenamefont {Hurand}\ \emph {et~al.}(2015)\citenamefont {Hurand},
  \citenamefont {Jouan}, \citenamefont {Feuillet-Palma}, \citenamefont {Singh},
  \citenamefont {Biscaras}, \citenamefont {Lesne}, \citenamefont {Reyren},
  \citenamefont {Barth\'{e}l\'{e}my}, \citenamefont {Bibes}, \citenamefont
  {Villegas}, \citenamefont {Ulysse}, \citenamefont {Lafosse}, \citenamefont
  {Pannetier-Lecoeur}, \citenamefont {Caprara}, \citenamefont {Grilli},
  \citenamefont {Lesueur},\ and\ \citenamefont {Bergeal}}]{RN407}%
  \BibitemOpen
  \bibfield  {author} {\bibinfo {author} {\bibfnamefont {S.}~\bibnamefont
  {Hurand}}, \bibinfo {author} {\bibfnamefont {A.}~\bibnamefont {Jouan}},
  \bibinfo {author} {\bibfnamefont {C.}~\bibnamefont {Feuillet-Palma}},
  \bibinfo {author} {\bibfnamefont {G.}~\bibnamefont {Singh}}, \bibinfo
  {author} {\bibfnamefont {J.}~\bibnamefont {Biscaras}}, \bibinfo {author}
  {\bibfnamefont {E.}~\bibnamefont {Lesne}}, \bibinfo {author} {\bibfnamefont
  {N.}~\bibnamefont {Reyren}}, \bibinfo {author} {\bibfnamefont
  {A.}~\bibnamefont {Barth\'{e}l\'{e}my}}, \bibinfo {author} {\bibfnamefont
  {M.}~\bibnamefont {Bibes}}, \bibinfo {author} {\bibfnamefont {J.~E.}\
  \bibnamefont {Villegas}}, \bibinfo {author} {\bibfnamefont {C.}~\bibnamefont
  {Ulysse}}, \bibinfo {author} {\bibfnamefont {X.}~\bibnamefont {Lafosse}},
  \bibinfo {author} {\bibfnamefont {M.}~\bibnamefont {Pannetier-Lecoeur}},
  \bibinfo {author} {\bibfnamefont {S.}~\bibnamefont {Caprara}}, \bibinfo
  {author} {\bibfnamefont {M.}~\bibnamefont {Grilli}}, \bibinfo {author}
  {\bibfnamefont {J.}~\bibnamefont {Lesueur}}, \ and\ \bibinfo {author}
  {\bibfnamefont {N.}~\bibnamefont {Bergeal}},\ }\href {\doibase
  10.1038/srep12751
  https://www.nature.com/articles/srep12751#supplementary-information}
  {\bibfield  {journal} {\bibinfo  {journal} {Sci. Rep.}\ }\textbf {\bibinfo
  {volume} {5}},\ \bibinfo {pages} {12751} (\bibinfo {year}
  {2015})}\BibitemShut {NoStop}%
\bibitem [{\citenamefont {Liang}\ \emph {et~al.}(2015)\citenamefont {Liang},
  \citenamefont {Cheng}, \citenamefont {Wei}, \citenamefont {Luo},
  \citenamefont {Yu}, \citenamefont {Zeng},\ and\ \citenamefont
  {Zhang}}]{RN372}%
  \BibitemOpen
  \bibfield  {author} {\bibinfo {author} {\bibfnamefont {H.}~\bibnamefont
  {Liang}}, \bibinfo {author} {\bibfnamefont {L.}~\bibnamefont {Cheng}},
  \bibinfo {author} {\bibfnamefont {L.}~\bibnamefont {Wei}}, \bibinfo {author}
  {\bibfnamefont {Z.}~\bibnamefont {Luo}}, \bibinfo {author} {\bibfnamefont
  {G.}~\bibnamefont {Yu}}, \bibinfo {author} {\bibfnamefont {C.}~\bibnamefont
  {Zeng}}, \ and\ \bibinfo {author} {\bibfnamefont {Z.}~\bibnamefont {Zhang}},\
  }\href {http://link.aps.org/doi/10.1103/PhysRevB.92.075309} {\bibfield
  {journal} {\bibinfo  {journal} {Phys. Rev. B}\ }\textbf {\bibinfo {volume}
  {92}},\ \bibinfo {pages} {075309} (\bibinfo {year} {2015})}\BibitemShut
  {NoStop}%
\bibitem [{\citenamefont {Santander-Syro}\ \emph {et~al.}(2014)\citenamefont
  {Santander-Syro}, \citenamefont {Fortuna}, \citenamefont {Bareille},
  \citenamefont {Rödel}, \citenamefont {Landolt}, \citenamefont {Plumb},
  \citenamefont {Dil},\ and\ \citenamefont {Radović}}]{RN415}%
  \BibitemOpen
  \bibfield  {author} {\bibinfo {author} {\bibfnamefont {A.~F.}\ \bibnamefont
  {Santander-Syro}}, \bibinfo {author} {\bibfnamefont {F.}~\bibnamefont
  {Fortuna}}, \bibinfo {author} {\bibfnamefont {C.}~\bibnamefont {Bareille}},
  \bibinfo {author} {\bibfnamefont {T.~C.}\ \bibnamefont {Rödel}}, \bibinfo
  {author} {\bibfnamefont {G.}~\bibnamefont {Landolt}}, \bibinfo {author}
  {\bibfnamefont {N.~C.}\ \bibnamefont {Plumb}}, \bibinfo {author}
  {\bibfnamefont {J.~H.}\ \bibnamefont {Dil}}, \ and\ \bibinfo {author}
  {\bibfnamefont {M.}~\bibnamefont {Radović}},\ }\href {\doibase
  10.1038/nmat4107
  https://www.nature.com/articles/nmat4107#supplementary-information}
  {\bibfield  {journal} {\bibinfo  {journal} {Nat. Mater.}\ }\textbf {\bibinfo
  {volume} {13}},\ \bibinfo {pages} {1085} (\bibinfo {year}
  {2014})}\BibitemShut {NoStop}%
\bibitem [{\citenamefont {Nakamura}\ \emph {et~al.}(2012)\citenamefont
  {Nakamura}, \citenamefont {Koga},\ and\ \citenamefont {Kimura}}]{RN408}%
  \BibitemOpen
  \bibfield  {author} {\bibinfo {author} {\bibfnamefont {H.}~\bibnamefont
  {Nakamura}}, \bibinfo {author} {\bibfnamefont {T.}~\bibnamefont {Koga}}, \
  and\ \bibinfo {author} {\bibfnamefont {T.}~\bibnamefont {Kimura}},\ }\href
  {https://link.aps.org/doi/10.1103/PhysRevLett.108.206601} {\bibfield
  {journal} {\bibinfo  {journal} {Phys. Rev. Lett.}\ }\textbf {\bibinfo
  {volume} {108}},\ \bibinfo {pages} {206601} (\bibinfo {year}
  {2012})}\BibitemShut {NoStop}%
\bibitem [{\citenamefont {Herranz}\ \emph {et~al.}(2015)\citenamefont
  {Herranz}, \citenamefont {Singh}, \citenamefont {Bergeal}, \citenamefont
  {Jouan}, \citenamefont {Lesueur}, \citenamefont {Gázquez}, \citenamefont
  {Varela}, \citenamefont {Scigaj}, \citenamefont {Dix}, \citenamefont
  {Sánchez},\ and\ \citenamefont {Fontcuberta}}]{RN365}%
  \BibitemOpen
  \bibfield  {author} {\bibinfo {author} {\bibfnamefont {G.}~\bibnamefont
  {Herranz}}, \bibinfo {author} {\bibfnamefont {G.}~\bibnamefont {Singh}},
  \bibinfo {author} {\bibfnamefont {N.}~\bibnamefont {Bergeal}}, \bibinfo
  {author} {\bibfnamefont {A.}~\bibnamefont {Jouan}}, \bibinfo {author}
  {\bibfnamefont {J.}~\bibnamefont {Lesueur}}, \bibinfo {author} {\bibfnamefont
  {J.}~\bibnamefont {Gázquez}}, \bibinfo {author} {\bibfnamefont
  {M.}~\bibnamefont {Varela}}, \bibinfo {author} {\bibfnamefont
  {M.}~\bibnamefont {Scigaj}}, \bibinfo {author} {\bibfnamefont
  {N.}~\bibnamefont {Dix}}, \bibinfo {author} {\bibfnamefont {F.}~\bibnamefont
  {Sánchez}}, \ and\ \bibinfo {author} {\bibfnamefont {J.}~\bibnamefont
  {Fontcuberta}},\ }\href {\doibase 10.1038/ncomms7028
  http://www.nature.com/articles/ncomms7028#supplementary-information}
  {\bibfield  {journal} {\bibinfo  {journal} {Nat. Commun.}\ }\textbf {\bibinfo
  {volume} {6}},\ \bibinfo {pages} {6028} (\bibinfo {year} {2015})}\BibitemShut
  {NoStop}%
\bibitem [{\citenamefont {Pesquera}\ \emph {et~al.}(2014)\citenamefont
  {Pesquera}, \citenamefont {Scigaj}, \citenamefont {Gargiani}, \citenamefont
  {Barla}, \citenamefont {Herrero-Mart\'{i}n}, \citenamefont {Pellegrin},
  \citenamefont {Valvidares}, \citenamefont {G\'{a}zquez}, \citenamefont
  {Varela}, \citenamefont {Dix}, \citenamefont {Fontcuberta}, \citenamefont
  {S\'{a}nchez},\ and\ \citenamefont {Herranz}}]{RN369}%
  \BibitemOpen
  \bibfield  {author} {\bibinfo {author} {\bibfnamefont {D.}~\bibnamefont
  {Pesquera}}, \bibinfo {author} {\bibfnamefont {M.}~\bibnamefont {Scigaj}},
  \bibinfo {author} {\bibfnamefont {P.}~\bibnamefont {Gargiani}}, \bibinfo
  {author} {\bibfnamefont {A.}~\bibnamefont {Barla}}, \bibinfo {author}
  {\bibfnamefont {J.}~\bibnamefont {Herrero-Mart\'{i}n}}, \bibinfo {author}
  {\bibfnamefont {E.}~\bibnamefont {Pellegrin}}, \bibinfo {author}
  {\bibfnamefont {S.~M.}\ \bibnamefont {Valvidares}}, \bibinfo {author}
  {\bibfnamefont {J.}~\bibnamefont {G\'{a}zquez}}, \bibinfo {author}
  {\bibfnamefont {M.}~\bibnamefont {Varela}}, \bibinfo {author} {\bibfnamefont
  {N.}~\bibnamefont {Dix}}, \bibinfo {author} {\bibfnamefont {J.}~\bibnamefont
  {Fontcuberta}}, \bibinfo {author} {\bibfnamefont {F.}~\bibnamefont
  {S\'{a}nchez}}, \ and\ \bibinfo {author} {\bibfnamefont {G.}~\bibnamefont
  {Herranz}},\ }\href {http://link.aps.org/doi/10.1103/PhysRevLett.113.156802}
  {\bibfield  {journal} {\bibinfo  {journal} {Phys. Rev. Lett.}\ }\textbf
  {\bibinfo {volume} {113}},\ \bibinfo {pages} {156802} (\bibinfo {year}
  {2014})}\BibitemShut {NoStop}%
\bibitem [{\citenamefont {Wang}\ \emph {et~al.}(2014)\citenamefont {Wang},
  \citenamefont {Zhong}, \citenamefont {Hao}, \citenamefont {Gerhold},
  \citenamefont {Stöger}, \citenamefont {Schmid}, \citenamefont
  {S\'{a}nchez-Barriga}, \citenamefont {Varykhalov}, \citenamefont {Franchini},
  \citenamefont {Held},\ and\ \citenamefont {Diebold}}]{RN370}%
  \BibitemOpen
  \bibfield  {author} {\bibinfo {author} {\bibfnamefont {Z.}~\bibnamefont
  {Wang}}, \bibinfo {author} {\bibfnamefont {Z.}~\bibnamefont {Zhong}},
  \bibinfo {author} {\bibfnamefont {X.}~\bibnamefont {Hao}}, \bibinfo {author}
  {\bibfnamefont {S.}~\bibnamefont {Gerhold}}, \bibinfo {author} {\bibfnamefont
  {B.}~\bibnamefont {Stöger}}, \bibinfo {author} {\bibfnamefont
  {M.}~\bibnamefont {Schmid}}, \bibinfo {author} {\bibfnamefont
  {J.}~\bibnamefont {S\'{a}nchez-Barriga}}, \bibinfo {author} {\bibfnamefont
  {A.}~\bibnamefont {Varykhalov}}, \bibinfo {author} {\bibfnamefont
  {C.}~\bibnamefont {Franchini}}, \bibinfo {author} {\bibfnamefont
  {K.}~\bibnamefont {Held}}, \ and\ \bibinfo {author} {\bibfnamefont
  {U.}~\bibnamefont {Diebold}},\ }\href {\doibase 10.1073/pnas.1318304111}
  {\bibfield  {journal} {\bibinfo  {journal} {PNAS}\ }\textbf {\bibinfo
  {volume} {111}},\ \bibinfo {pages} {3933} (\bibinfo {year}
  {2014})}\BibitemShut {NoStop}%
\bibitem [{\citenamefont {Winkler}(2003)}]{winkler2003spin}%
  \BibitemOpen
  \bibfield  {author} {\bibinfo {author} {\bibfnamefont {R.}~\bibnamefont
  {Winkler}},\ }\href@noop {} {\emph {\bibinfo {title} {Spin-orbit coupling
  effects in two-dimensional electron and hole systems}}},\ Vol.\ \bibinfo
  {volume} {191}\ (\bibinfo  {publisher} {Springer Science \& Business Media},\
  \bibinfo {year} {2003})\BibitemShut {NoStop}%
\bibitem [{\citenamefont {Song}\ \emph {et~al.}(2017)\citenamefont {Song},
  \citenamefont {Zhang}, \citenamefont {Su}, \citenamefont {Yuan},
  \citenamefont {Chen}, \citenamefont {Xing}, \citenamefont {Shi},
  \citenamefont {Sun},\ and\ \citenamefont {Han}}]{RN412}%
  \BibitemOpen
  \bibfield  {author} {\bibinfo {author} {\bibfnamefont {Q.}~\bibnamefont
  {Song}}, \bibinfo {author} {\bibfnamefont {H.}~\bibnamefont {Zhang}},
  \bibinfo {author} {\bibfnamefont {T.}~\bibnamefont {Su}}, \bibinfo {author}
  {\bibfnamefont {W.}~\bibnamefont {Yuan}}, \bibinfo {author} {\bibfnamefont
  {Y.}~\bibnamefont {Chen}}, \bibinfo {author} {\bibfnamefont {W.}~\bibnamefont
  {Xing}}, \bibinfo {author} {\bibfnamefont {J.}~\bibnamefont {Shi}}, \bibinfo
  {author} {\bibfnamefont {J.}~\bibnamefont {Sun}}, \ and\ \bibinfo {author}
  {\bibfnamefont {W.}~\bibnamefont {Han}},\ }\href {\doibase
  10.1126/sciadv.1602312} {\bibfield  {journal} {\bibinfo  {journal} {Sci.
  Adv.}\ }\textbf {\bibinfo {volume} {3}},\ \bibinfo {pages} {e1602312}
  (\bibinfo {year} {2017})}\BibitemShut {NoStop}%
\bibitem [{\citenamefont {Jin}\ \emph {et~al.}(2016)\citenamefont {Jin},
  \citenamefont {Moon}, \citenamefont {Park}, \citenamefont {Modepalli},
  \citenamefont {Jo}, \citenamefont {Kim}, \citenamefont {Koo}, \citenamefont
  {Min}, \citenamefont {Lee}, \citenamefont {Baek} \emph
  {et~al.}}]{jin2016nonlocal}%
  \BibitemOpen
  \bibfield  {author} {\bibinfo {author} {\bibfnamefont {M.-J.}\ \bibnamefont
  {Jin}}, \bibinfo {author} {\bibfnamefont {S.~Y.}\ \bibnamefont {Moon}},
  \bibinfo {author} {\bibfnamefont {J.}~\bibnamefont {Park}}, \bibinfo {author}
  {\bibfnamefont {V.}~\bibnamefont {Modepalli}}, \bibinfo {author}
  {\bibfnamefont {J.}~\bibnamefont {Jo}}, \bibinfo {author} {\bibfnamefont
  {S.-I.}\ \bibnamefont {Kim}}, \bibinfo {author} {\bibfnamefont {H.~C.}\
  \bibnamefont {Koo}}, \bibinfo {author} {\bibfnamefont {B.-C.}\ \bibnamefont
  {Min}}, \bibinfo {author} {\bibfnamefont {H.-W.}\ \bibnamefont {Lee}},
  \bibinfo {author} {\bibfnamefont {S.-H.}\ \bibnamefont {Baek}},  \emph
  {et~al.},\ }\href@noop {} {\bibfield  {journal} {\bibinfo  {journal} {Nano
  letters}\ }\textbf {\bibinfo {volume} {17}},\ \bibinfo {pages} {36} (\bibinfo
  {year} {2016})}\BibitemShut {NoStop}%
\bibitem [{\citenamefont {Chauleau}\ \emph {et~al.}(2016)\citenamefont
  {Chauleau}, \citenamefont {Boselli}, \citenamefont {Gariglio}, \citenamefont
  {Weil}, \citenamefont {Loubens}, \citenamefont {Triscone},\ and\
  \citenamefont {Viret}}]{RN413}%
  \BibitemOpen
  \bibfield  {author} {\bibinfo {author} {\bibfnamefont {J.~Y.}\ \bibnamefont
  {Chauleau}}, \bibinfo {author} {\bibfnamefont {M.}~\bibnamefont {Boselli}},
  \bibinfo {author} {\bibfnamefont {S.}~\bibnamefont {Gariglio}}, \bibinfo
  {author} {\bibfnamefont {R.}~\bibnamefont {Weil}}, \bibinfo {author}
  {\bibfnamefont {G.~d.}\ \bibnamefont {Loubens}}, \bibinfo {author}
  {\bibfnamefont {J.~M.}\ \bibnamefont {Triscone}}, \ and\ \bibinfo {author}
  {\bibfnamefont {M.}~\bibnamefont {Viret}},\ }\href
  {http://stacks.iop.org/0295-5075/116/i=1/a=17006} {\bibfield  {journal}
  {\bibinfo  {journal} {EPL (EuroPhysics Letters)}\ }\textbf {\bibinfo {volume}
  {116}},\ \bibinfo {pages} {17006} (\bibinfo {year} {2016})}\BibitemShut
  {NoStop}%
\bibitem [{\citenamefont {Lesne}\ \emph {et~al.}(2016)\citenamefont {Lesne},
  \citenamefont {Fu}, \citenamefont {Oyarzun}, \citenamefont
  {Rojas-S\'{a}nchez}, \citenamefont {Vaz}, \citenamefont {Naganuma},
  \citenamefont {Sicoli}, \citenamefont {Attan\'{e}}, \citenamefont {Jamet},
  \citenamefont {Jacquet}, \citenamefont {George}, \citenamefont
  {Barth\'{e}l\'{e}my}, \citenamefont {Jaffrès}, \citenamefont {Fert},
  \citenamefont {Bibes},\ and\ \citenamefont {Vila}}]{RN414}%
  \BibitemOpen
  \bibfield  {author} {\bibinfo {author} {\bibfnamefont {E.}~\bibnamefont
  {Lesne}}, \bibinfo {author} {\bibfnamefont {Y.}~\bibnamefont {Fu}}, \bibinfo
  {author} {\bibfnamefont {S.}~\bibnamefont {Oyarzun}}, \bibinfo {author}
  {\bibfnamefont {J.~C.}\ \bibnamefont {Rojas-S\'{a}nchez}}, \bibinfo {author}
  {\bibfnamefont {D.~C.}\ \bibnamefont {Vaz}}, \bibinfo {author} {\bibfnamefont
  {H.}~\bibnamefont {Naganuma}}, \bibinfo {author} {\bibfnamefont
  {G.}~\bibnamefont {Sicoli}}, \bibinfo {author} {\bibfnamefont {J.~P.}\
  \bibnamefont {Attan\'{e}}}, \bibinfo {author} {\bibfnamefont
  {M.}~\bibnamefont {Jamet}}, \bibinfo {author} {\bibfnamefont
  {E.}~\bibnamefont {Jacquet}}, \bibinfo {author} {\bibfnamefont {J.~M.}\
  \bibnamefont {George}}, \bibinfo {author} {\bibfnamefont {A.}~\bibnamefont
  {Barth\'{e}l\'{e}my}}, \bibinfo {author} {\bibfnamefont {H.}~\bibnamefont
  {Jaffrès}}, \bibinfo {author} {\bibfnamefont {A.}~\bibnamefont {Fert}},
  \bibinfo {author} {\bibfnamefont {M.}~\bibnamefont {Bibes}}, \ and\ \bibinfo
  {author} {\bibfnamefont {L.}~\bibnamefont {Vila}},\ }\href {\doibase
  10.1038/nmat4726
  https://www.nature.com/articles/nmat4726#supplementary-information}
  {\bibfield  {journal} {\bibinfo  {journal} {Nat. Mater.}\ }\textbf {\bibinfo
  {volume} {15}},\ \bibinfo {pages} {1261} (\bibinfo {year}
  {2016})}\BibitemShut {NoStop}%
\bibitem [{\citenamefont {Narayanapillai}\ \emph {et~al.}(2014)\citenamefont
  {Narayanapillai}, \citenamefont {Gopinadhan}, \citenamefont {Qiu},
  \citenamefont {Annadi}, \citenamefont {Ariando}, \citenamefont {Venkatesan},\
  and\ \citenamefont {Yang}}]{RN418}%
  \BibitemOpen
  \bibfield  {author} {\bibinfo {author} {\bibfnamefont {K.}~\bibnamefont
  {Narayanapillai}}, \bibinfo {author} {\bibfnamefont {K.}~\bibnamefont
  {Gopinadhan}}, \bibinfo {author} {\bibfnamefont {X.}~\bibnamefont {Qiu}},
  \bibinfo {author} {\bibfnamefont {A.}~\bibnamefont {Annadi}}, \bibinfo
  {author} {\bibnamefont {Ariando}}, \bibinfo {author} {\bibfnamefont
  {T.}~\bibnamefont {Venkatesan}}, \ and\ \bibinfo {author} {\bibfnamefont
  {H.}~\bibnamefont {Yang}},\ }\href {\doibase 10.1063/1.4899122} {\bibfield
  {journal} {\bibinfo  {journal} {Appl. Phys. Lett.}\ }\textbf {\bibinfo
  {volume} {105}},\ \bibinfo {pages} {162405} (\bibinfo {year}
  {2014})}\BibitemShut {NoStop}%
\bibitem [{\citenamefont {Wang}\ \emph {et~al.}(2017)\citenamefont {Wang},
  \citenamefont {Ramaswamy}, \citenamefont {Motapothula}, \citenamefont
  {Narayanapillai}, \citenamefont {Zhu}, \citenamefont {Yu}, \citenamefont
  {Venkatesan},\ and\ \citenamefont {Yang}}]{RN411}%
  \BibitemOpen
  \bibfield  {author} {\bibinfo {author} {\bibfnamefont {Y.}~\bibnamefont
  {Wang}}, \bibinfo {author} {\bibfnamefont {R.}~\bibnamefont {Ramaswamy}},
  \bibinfo {author} {\bibfnamefont {M.}~\bibnamefont {Motapothula}}, \bibinfo
  {author} {\bibfnamefont {K.}~\bibnamefont {Narayanapillai}}, \bibinfo
  {author} {\bibfnamefont {D.}~\bibnamefont {Zhu}}, \bibinfo {author}
  {\bibfnamefont {J.}~\bibnamefont {Yu}}, \bibinfo {author} {\bibfnamefont
  {T.}~\bibnamefont {Venkatesan}}, \ and\ \bibinfo {author} {\bibfnamefont
  {H.}~\bibnamefont {Yang}},\ }\href {\doibase 10.1021/acs.nanolett.7b03714}
  {\bibfield  {journal} {\bibinfo  {journal} {Nano Lett.}\ }\textbf {\bibinfo
  {volume} {17}},\ \bibinfo {pages} {7659} (\bibinfo {year}
  {2017})}\BibitemShut {NoStop}%
\bibitem [{\citenamefont {Huang}\ \emph
  {et~al.}(2018{\natexlab{b}})\citenamefont {Huang}, \citenamefont
  {Renshaw~Wang}, \citenamefont {Rusydi}, \citenamefont {Chen}, \citenamefont
  {Yang},\ and\ \citenamefont {Venkatesan}}]{huang2018interface}%
  \BibitemOpen
  \bibfield  {author} {\bibinfo {author} {\bibfnamefont {Z.}~\bibnamefont
  {Huang}}, \bibinfo {author} {\bibfnamefont {X.}~\bibnamefont {Renshaw~Wang}},
  \bibinfo {author} {\bibfnamefont {A.}~\bibnamefont {Rusydi}}, \bibinfo
  {author} {\bibfnamefont {J.}~\bibnamefont {Chen}}, \bibinfo {author}
  {\bibfnamefont {H.}~\bibnamefont {Yang}}, \ and\ \bibinfo {author}
  {\bibfnamefont {T.}~\bibnamefont {Venkatesan}},\ }\href@noop {} {\bibfield
  {journal} {\bibinfo  {journal} {Advanced Materials}\ }\textbf {\bibinfo
  {volume} {30}},\ \bibinfo {pages} {1802439} (\bibinfo {year}
  {2018}{\natexlab{b}})}\BibitemShut {NoStop}%
\bibitem [{\citenamefont {Pai}\ \emph {et~al.}(2018)\citenamefont {Pai},
  \citenamefont {Tylan-Tyler}, \citenamefont {Irvin},\ and\ \citenamefont
  {Levy}}]{pai2018physics}%
  \BibitemOpen
  \bibfield  {author} {\bibinfo {author} {\bibfnamefont {Y.-Y.}\ \bibnamefont
  {Pai}}, \bibinfo {author} {\bibfnamefont {A.}~\bibnamefont {Tylan-Tyler}},
  \bibinfo {author} {\bibfnamefont {P.}~\bibnamefont {Irvin}}, \ and\ \bibinfo
  {author} {\bibfnamefont {J.}~\bibnamefont {Levy}},\ }\href@noop {} {\bibfield
   {journal} {\bibinfo  {journal} {Reports on Progress in Physics}\ }\textbf
  {\bibinfo {volume} {81}},\ \bibinfo {pages} {036503} (\bibinfo {year}
  {2018})}\BibitemShut {NoStop}%
\bibitem [{\citenamefont {Bychkov}\ and\ \citenamefont
  {Rashba}(1984)}]{Rashba1984}%
  \BibitemOpen
  \bibfield  {author} {\bibinfo {author} {\bibfnamefont {Y.~A.}\ \bibnamefont
  {Bychkov}}\ and\ \bibinfo {author} {\bibfnamefont {E.~I.}\ \bibnamefont
  {Rashba}},\ }\href {http://stacks.iop.org/0022-3719/17/i=33/a=015} {\bibfield
   {journal} {\bibinfo  {journal} {J. Phys. C: Solid State Phys.}\ }\textbf
  {\bibinfo {volume} {17}},\ \bibinfo {pages} {6039} (\bibinfo {year}
  {1984})}\BibitemShut {NoStop}%
\bibitem [{\citenamefont {Khalsa}\ \emph {et~al.}(2013)\citenamefont {Khalsa},
  \citenamefont {Lee},\ and\ \citenamefont {MacDonald}}]{PhysRevB.88.041302}%
  \BibitemOpen
  \bibfield  {author} {\bibinfo {author} {\bibfnamefont {G.}~\bibnamefont
  {Khalsa}}, \bibinfo {author} {\bibfnamefont {B.}~\bibnamefont {Lee}}, \ and\
  \bibinfo {author} {\bibfnamefont {A.~H.}\ \bibnamefont {MacDonald}},\ }\href
  {\doibase 10.1103/PhysRevB.88.041302} {\bibfield  {journal} {\bibinfo
  {journal} {Phys. Rev. B}\ }\textbf {\bibinfo {volume} {88}},\ \bibinfo
  {pages} {041302} (\bibinfo {year} {2013})}\BibitemShut {NoStop}%
\bibitem [{\citenamefont {Zhou}\ \emph {et~al.}(2015)\citenamefont {Zhou},
  \citenamefont {Shan},\ and\ \citenamefont {Xiao}}]{Zhou:PRB15}%
  \BibitemOpen
  \bibfield  {author} {\bibinfo {author} {\bibfnamefont {J.}~\bibnamefont
  {Zhou}}, \bibinfo {author} {\bibfnamefont {W.-Y.}\ \bibnamefont {Shan}}, \
  and\ \bibinfo {author} {\bibfnamefont {D.}~\bibnamefont {Xiao}},\ }\href
  {\doibase 10.1103/PhysRevB.91.241302} {\bibfield  {journal} {\bibinfo
  {journal} {Phys. Rev. B}\ }\textbf {\bibinfo {volume} {91}},\ \bibinfo
  {pages} {241302} (\bibinfo {year} {2015})}\BibitemShut {NoStop}%
\bibitem [{\citenamefont {van Heeringen}\ \emph {et~al.}(2013)\citenamefont
  {van Heeringen}, \citenamefont {de~Wijs}, \citenamefont {McCollam},
  \citenamefont {Maan},\ and\ \citenamefont {Fasolino}}]{RN401}%
  \BibitemOpen
  \bibfield  {author} {\bibinfo {author} {\bibfnamefont {L.~W.}\ \bibnamefont
  {van Heeringen}}, \bibinfo {author} {\bibfnamefont {G.~A.}\ \bibnamefont
  {de~Wijs}}, \bibinfo {author} {\bibfnamefont {A.}~\bibnamefont {McCollam}},
  \bibinfo {author} {\bibfnamefont {J.~C.}\ \bibnamefont {Maan}}, \ and\
  \bibinfo {author} {\bibfnamefont {A.}~\bibnamefont {Fasolino}},\ }\href
  {https://link.aps.org/doi/10.1103/PhysRevB.88.205140} {\bibfield  {journal}
  {\bibinfo  {journal} {Phys. Rev. B}\ }\textbf {\bibinfo {volume} {88}},\
  \bibinfo {pages} {205140} (\bibinfo {year} {2013})}\BibitemShut {NoStop}%
\bibitem [{\citenamefont {van Heeringen}\ \emph {et~al.}(2017)\citenamefont
  {van Heeringen}, \citenamefont {McCollam}, \citenamefont {de~Wijs},\ and\
  \citenamefont {Fasolino}}]{k.p2}%
  \BibitemOpen
  \bibfield  {author} {\bibinfo {author} {\bibfnamefont {L.~W.}\ \bibnamefont
  {van Heeringen}}, \bibinfo {author} {\bibfnamefont {A.}~\bibnamefont
  {McCollam}}, \bibinfo {author} {\bibfnamefont {G.~A.}\ \bibnamefont
  {de~Wijs}}, \ and\ \bibinfo {author} {\bibfnamefont {A.}~\bibnamefont
  {Fasolino}},\ }\href {\doibase 10.1103/PhysRevB.95.155134} {\bibfield
  {journal} {\bibinfo  {journal} {Phys. Rev. B}\ }\textbf {\bibinfo {volume}
  {95}},\ \bibinfo {pages} {155134} (\bibinfo {year} {2017})}\BibitemShut
  {NoStop}%
\bibitem [{\citenamefont {Ohtomo}\ and\ \citenamefont
  {Hwang}(2004{\natexlab{b}})}]{RN358}%
  \BibitemOpen
  \bibfield  {author} {\bibinfo {author} {\bibfnamefont {A.}~\bibnamefont
  {Ohtomo}}\ and\ \bibinfo {author} {\bibfnamefont {H.~Y.}\ \bibnamefont
  {Hwang}},\ }\href {http://dx.doi.org/10.1038/nature02308} {\bibfield
  {journal} {\bibinfo  {journal} {Nature}\ }\textbf {\bibinfo {volume} {427}},\
  \bibinfo {pages} {423} (\bibinfo {year} {2004}{\natexlab{b}})}\BibitemShut
  {NoStop}%
\bibitem [{\citenamefont {Bistritzer}\ \emph {et~al.}(2011)\citenamefont
  {Bistritzer}, \citenamefont {Khalsa},\ and\ \citenamefont
  {MacDonald}}]{RN400}%
  \BibitemOpen
  \bibfield  {author} {\bibinfo {author} {\bibfnamefont {R.}~\bibnamefont
  {Bistritzer}}, \bibinfo {author} {\bibfnamefont {G.}~\bibnamefont {Khalsa}},
  \ and\ \bibinfo {author} {\bibfnamefont {A.~H.}\ \bibnamefont {MacDonald}},\
  }\href {https://link.aps.org/doi/10.1103/PhysRevB.83.115114} {\bibfield
  {journal} {\bibinfo  {journal} {Phys. Rev. B}\ }\textbf {\bibinfo {volume}
  {83}},\ \bibinfo {pages} {115114} (\bibinfo {year} {2011})}\BibitemShut
  {NoStop}%
\bibitem [{\citenamefont {Khalsa}\ and\ \citenamefont
  {MacDonald}(2012)}]{tetragonal:PRB2012}%
  \BibitemOpen
  \bibfield  {author} {\bibinfo {author} {\bibfnamefont {G.}~\bibnamefont
  {Khalsa}}\ and\ \bibinfo {author} {\bibfnamefont {A.~H.}\ \bibnamefont
  {MacDonald}},\ }\href {\doibase 10.1103/PhysRevB.86.125121} {\bibfield
  {journal} {\bibinfo  {journal} {Phys. Rev. B}\ }\textbf {\bibinfo {volume}
  {86}},\ \bibinfo {pages} {125121} (\bibinfo {year} {2012})}\BibitemShut
  {NoStop}%
\bibitem [{\citenamefont {Kim}\ \emph {et~al.}(2013)\citenamefont {Kim},
  \citenamefont {Lutchyn},\ and\ \citenamefont {Nayak}}]{Kim:PRB13}%
  \BibitemOpen
  \bibfield  {author} {\bibinfo {author} {\bibfnamefont {Y.}~\bibnamefont
  {Kim}}, \bibinfo {author} {\bibfnamefont {R.~M.}\ \bibnamefont {Lutchyn}}, \
  and\ \bibinfo {author} {\bibfnamefont {C.}~\bibnamefont {Nayak}},\ }\href
  {\doibase 10.1103/PhysRevB.87.245121} {\bibfield  {journal} {\bibinfo
  {journal} {Phys. Rev. B}\ }\textbf {\bibinfo {volume} {87}},\ \bibinfo
  {pages} {245121} (\bibinfo {year} {2013})}\BibitemShut {NoStop}%
\bibitem [{\citenamefont {LaShell}\ \emph {et~al.}(1996)\citenamefont
  {LaShell}, \citenamefont {McDougall},\ and\ \citenamefont {Jensen}}]{RN402}%
  \BibitemOpen
  \bibfield  {author} {\bibinfo {author} {\bibfnamefont {S.}~\bibnamefont
  {LaShell}}, \bibinfo {author} {\bibfnamefont {B.~A.}\ \bibnamefont
  {McDougall}}, \ and\ \bibinfo {author} {\bibfnamefont {E.}~\bibnamefont
  {Jensen}},\ }\href {https://link.aps.org/doi/10.1103/PhysRevLett.77.3419}
  {\bibfield  {journal} {\bibinfo  {journal} {Phys. Rev. Lett.}\ }\textbf
  {\bibinfo {volume} {77}},\ \bibinfo {pages} {3419} (\bibinfo {year}
  {1996})}\BibitemShut {NoStop}%
\bibitem [{\citenamefont {Petersen}\ and\ \citenamefont
  {Hedeg\r{a}rd}(2000)}]{RN403}%
  \BibitemOpen
  \bibfield  {author} {\bibinfo {author} {\bibfnamefont {L.}~\bibnamefont
  {Petersen}}\ and\ \bibinfo {author} {\bibfnamefont {P.}~\bibnamefont
  {Hedeg\r{a}rd}},\ }\href {\doibase
  https://doi.org/10.1016/S0039-6028(00)00441-6} {\bibfield  {journal}
  {\bibinfo  {journal} {Surf. Sci.}\ }\textbf {\bibinfo {volume} {459}},\
  \bibinfo {pages} {49} (\bibinfo {year} {2000})}\BibitemShut {NoStop}%
\bibitem [{\citenamefont {Reich}\ \emph {et~al.}(2015)\citenamefont {Reich},
  \citenamefont {Schecter},\ and\ \citenamefont {Shklovskii}}]{STOdensity}%
  \BibitemOpen
  \bibfield  {author} {\bibinfo {author} {\bibfnamefont {K.~V.}\ \bibnamefont
  {Reich}}, \bibinfo {author} {\bibfnamefont {M.}~\bibnamefont {Schecter}}, \
  and\ \bibinfo {author} {\bibfnamefont {B.~I.}\ \bibnamefont {Shklovskii}},\
  }\href {\doibase 10.1103/PhysRevB.91.115303} {\bibfield  {journal} {\bibinfo
  {journal} {Phys. Rev. B}\ }\textbf {\bibinfo {volume} {91}},\ \bibinfo
  {pages} {115303} (\bibinfo {year} {2015})}\BibitemShut {NoStop}%
\bibitem [{\citenamefont {Zhang}\ \emph {et~al.}(2018)\citenamefont {Zhang},
  \citenamefont {Pang}, \citenamefont {Chen},\ and\ \citenamefont
  {Chen}}]{SrIrO3:review}%
  \BibitemOpen
  \bibfield  {author} {\bibinfo {author} {\bibfnamefont {L.}~\bibnamefont
  {Zhang}}, \bibinfo {author} {\bibfnamefont {B.}~\bibnamefont {Pang}},
  \bibinfo {author} {\bibfnamefont {Y.~B.}\ \bibnamefont {Chen}}, \ and\
  \bibinfo {author} {\bibfnamefont {Y.}~\bibnamefont {Chen}},\ }\href {\doibase
  10.1080/10408436.2017.1358147} {\bibfield  {journal} {\bibinfo  {journal}
  {Critical Reviews in Solid State and Materials Sciences}\ }\textbf {\bibinfo
  {volume} {43}},\ \bibinfo {pages} {367} (\bibinfo {year} {2018})},\ \Eprint
  {http://arxiv.org/abs/https://doi.org/10.1080/10408436.2017.1358147}
  {https://doi.org/10.1080/10408436.2017.1358147} \BibitemShut {NoStop}%
\bibitem [{\citenamefont {Zhang}\ \emph {et~al.}(2014)\citenamefont {Zhang},
  \citenamefont {Chen}, \citenamefont {Zhang}, \citenamefont {Zhou},
  \citenamefont {Zhang}, \citenamefont {Gu}, \citenamefont {Yao},\ and\
  \citenamefont {Chen}}]{SrIrO3:RSOC}%
  \BibitemOpen
  \bibfield  {author} {\bibinfo {author} {\bibfnamefont {L.}~\bibnamefont
  {Zhang}}, \bibinfo {author} {\bibfnamefont {Y.}~\bibnamefont {Chen}},
  \bibinfo {author} {\bibfnamefont {B.}~\bibnamefont {Zhang}}, \bibinfo
  {author} {\bibfnamefont {J.}~\bibnamefont {Zhou}}, \bibinfo {author}
  {\bibfnamefont {S.}~\bibnamefont {Zhang}}, \bibinfo {author} {\bibfnamefont
  {Z.}~\bibnamefont {Gu}}, \bibinfo {author} {\bibfnamefont {S.}~\bibnamefont
  {Yao}}, \ and\ \bibinfo {author} {\bibfnamefont {Y.}~\bibnamefont {Chen}},\
  }\href@noop {} {\bibfield  {journal} {\bibinfo  {journal} {Journal of the
  Physical Society of Japan}\ }\textbf {\bibinfo {volume} {83}},\ \bibinfo
  {pages} {054707} (\bibinfo {year} {2014})}\BibitemShut {NoStop}%
\bibitem [{\citenamefont {Stengel}(2011)}]{orbital1}%
  \BibitemOpen
  \bibfield  {author} {\bibinfo {author} {\bibfnamefont {M.}~\bibnamefont
  {Stengel}},\ }\href {\doibase 10.1103/PhysRevLett.106.136803} {\bibfield
  {journal} {\bibinfo  {journal} {Phys. Rev. Lett.}\ }\textbf {\bibinfo
  {volume} {106}},\ \bibinfo {pages} {136803} (\bibinfo {year}
  {2011})}\BibitemShut {NoStop}%
\bibitem [{\citenamefont {Raslan}\ and\ \citenamefont
  {Atkinson}(2018)}]{orbital2}%
  \BibitemOpen
  \bibfield  {author} {\bibinfo {author} {\bibfnamefont {A.}~\bibnamefont
  {Raslan}}\ and\ \bibinfo {author} {\bibfnamefont {W.~A.}\ \bibnamefont
  {Atkinson}},\ }\href {\doibase 10.1103/PhysRevB.98.195447} {\bibfield
  {journal} {\bibinfo  {journal} {Phys. Rev. B}\ }\textbf {\bibinfo {volume}
  {98}},\ \bibinfo {pages} {195447} (\bibinfo {year} {2018})}\BibitemShut
  {NoStop}%
\bibitem [{\citenamefont {Bastard}\ \emph {et~al.}(1983)\citenamefont
  {Bastard}, \citenamefont {Mendez}, \citenamefont {Chang},\ and\ \citenamefont
  {Esaki}}]{Varmethod}%
  \BibitemOpen
  \bibfield  {author} {\bibinfo {author} {\bibfnamefont {G.}~\bibnamefont
  {Bastard}}, \bibinfo {author} {\bibfnamefont {E.~E.}\ \bibnamefont {Mendez}},
  \bibinfo {author} {\bibfnamefont {L.~L.}\ \bibnamefont {Chang}}, \ and\
  \bibinfo {author} {\bibfnamefont {L.}~\bibnamefont {Esaki}},\ }\href
  {\doibase 10.1103/PhysRevB.28.3241} {\bibfield  {journal} {\bibinfo
  {journal} {Phys. Rev. B}\ }\textbf {\bibinfo {volume} {28}},\ \bibinfo
  {pages} {3241} (\bibinfo {year} {1983})}\BibitemShut {NoStop}%
\end{thebibliography}
%

\pagebreak
\begin{widetext}

\appendix

\section{Determine $\bf{k\cdot p}$ parameters via DFT calculation}\label{S1}

To determine the $\bf{k\cdot p}$ parameters, first-principles calculations were performed using density-functional theory (DFT) based Vienna ab initio simulation package (VASP5.4.4.18) with local density approximation (LDA) for the exchange-correlation interaction. Projector augmented wave (PAW) potentials were selected to account for the interactions between electrons and ions. The LDA+U method was used with on-site effective U = 1.2 eV for the Ti d orbitals. The kinetic cut off energy for the expansion of plane-wave function was set to 500 eV. Monkhorst-Pack based k-point grids for sampling the first Brillouin-zone were set to $8\times 8\times 8$ and $8\times 8\times 1$ for bulk STO and LAO/STO respectively. The Hellman-Feynman forces on each atom were minimized with a tolerance value of $0.01 \mathrm{eV/\AA}$ for bulk STO, bulk LAO, and $\mathrm{(LAO)_{5.5}/(STO)_{30.5}}$ heterostructure. Based on these settings, the equilibrium lattice constant of the bulk STO is calculated $a=3.870 \mathrm{\AA}$, which is consistent with the experimental result (3.905 $\mathrm{\mathrm{\AA}}$).

{\it Without atomic SOC-} Along $k_{[100]}$, the dispersions are given by $Lk_{[100]}^2, Mk_{[100]}^2$. By fitting the calculated energies along $k_{[100]}$ direction to these $\bm{k\cdot p}$ formula [as shown in Fig. \ref{Fig1}a], we can obtain the values of $L=0.68  \ \mathrm{eV\AA^2}, M=9.23 \ \mathrm{eV\AA^2}$, respectively. Similarly, along $k_{[111]}$ the  dispersions are read as $1/3(L+2M-N)k_{[111]}^2,1/3(L+2M+2N)k_{[111]}^2$. The different between these two bands are $Nk_{[111]}^2$, and thus by fitting the dataset to this expression, we obtain $N=1.54  \ \mathrm{eV\AA^2}$.

{\it With SOC-} Similarly, in the presence of the atomic SOC, the three $t_{2g}$ bands are no longer degenerate at the $\Gamma$-point, and the gap is determined by the SOC splitting $\Delta_{SO}=29.85$ meV [as shown in Fig. \ref{Fig1}(b)]


\begin{figure}[h]
  \includegraphics[width=0.8\textwidth]{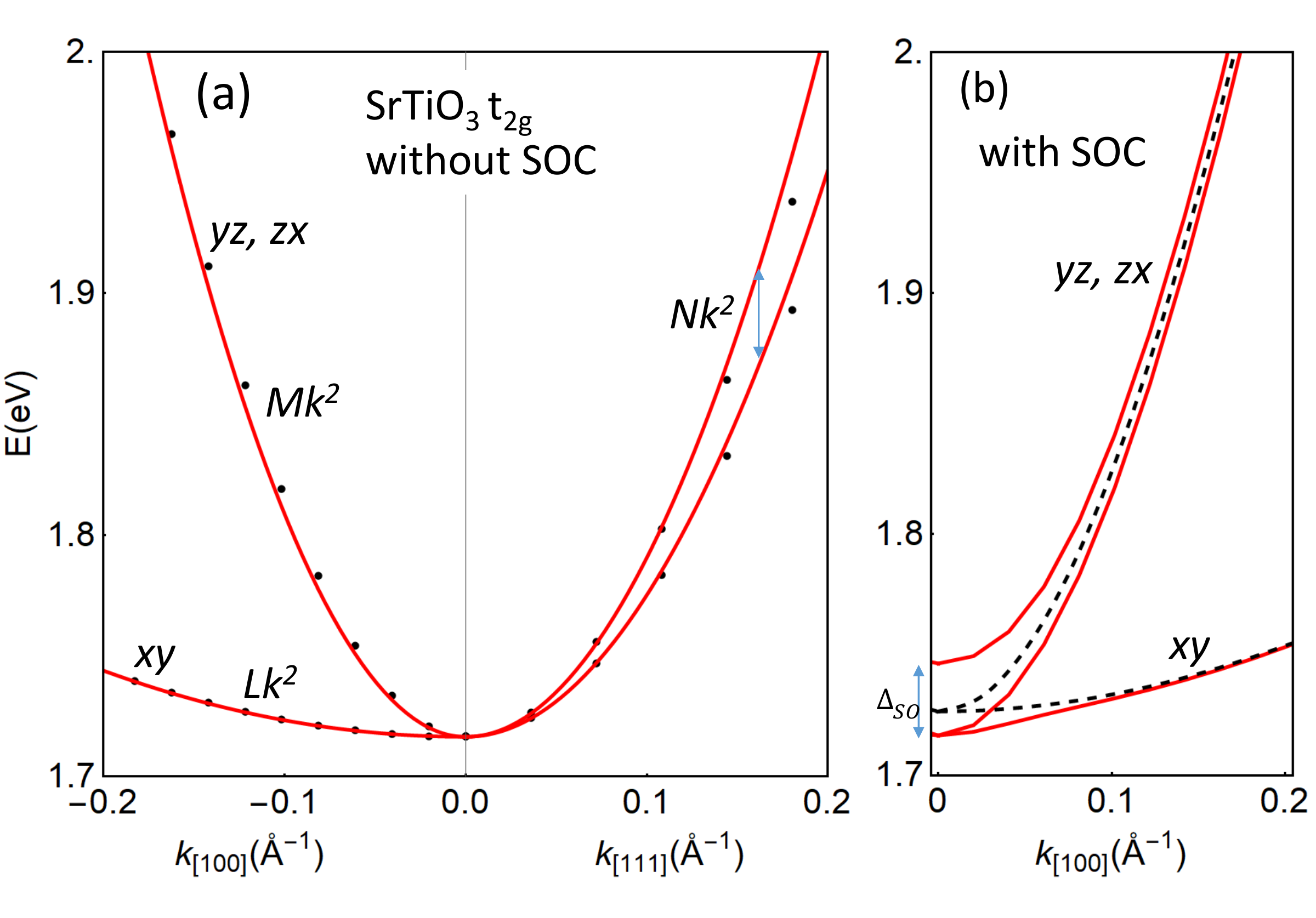}
  \caption{(a) Dispersion of the $t_{2g}$ bands in bulk SrTiO$_3$ obtained via DFT calculation without SOC. The fit of  the k.p model to the energy dispersion along [100] and [111] directions yields the mass parameters as $L=0.68\ \mathrm{eV\AA^2}, M=9.23 \ \mathrm{eV\AA^2}$, and $N=1.54  \ \mathrm{eV\AA^2}$, respectively. (b) Dispersion with SOC shows a splitting of $\Delta_{SO}=3\xi_{SO}=29.85$ meV.}
  \label{Fig1}
\end{figure}

\section{Eigenstates and eigenenergies of the $\Gamma-$ point Hamiltonian}\label{S2}

We introduce $|J,m_J\rangle$ basis given by
\begin{eqnarray}
u_1&&=|\frac{1}{2},\frac{1}{2}\rangle=\frac{1}{\sqrt{3}}\left(-i|yz,\drow\rangle+|zx,\drow\rangle-i|xy,\urow\rangle\right),\nonumber\\
u_2&&=|\frac{3}{2},\frac{1}{2}\rangle=\frac{1}{\sqrt{6}}\left(-i|yz,\drow\rangle+|zx,\drow\rangle+2i|xy,\urow\rangle\right),\nonumber\\
u_3&&=|\frac{3}{2},-\frac{1}{2}\rangle=\frac{1}{\sqrt{2}}\left(-i|yz,\drow\rangle-|zx,\drow\rangle\right),\nonumber\\
u_4&&=|\frac{3}{2},\frac{3}{2}\rangle=\frac{1}{\sqrt{3}}\left(-i|yz,\urow\rangle-|zx,\urow\rangle+i|xy,\drow\rangle\right),\nonumber\\
u_5&&=|\frac{1}{2},-\frac{1}{2}\rangle=\frac{1}{\sqrt{6}}\left(i|yz,\urow\rangle+|zx,\urow\rangle+2i|xy,\drow\rangle\right),\\
u_6&&=|\frac{3}{2},-\frac{3}{2}\rangle=\frac{1}{\sqrt{2}}\left(i|yz,\urow\rangle-|zx,\urow\rangle\right),\nonumber
\end{eqnarray}
in which the atomic SOC Hamiltonian is diagonalized $H_{SO}=\xi_{SO}{\mathrm{diag}}\left[2,-1,-1,2,-1,-1\right]$.   The  full Hamiltonian in this basis is given by

\begin{eqnarray}\label{Hfull}
H_\mathrm{full}=\begin{bmatrix}
h(q)&\Delta\\
\Delta^\dag &h(-q)^*
\end{bmatrix}.
\end{eqnarray}
where
\begin{eqnarray}
h(q)=
\begin{bmatrix}
a&q&r\\
q&b&r\\
r^*&r^*&c
\end{bmatrix},\ \
\Delta=\gamma \begin{bmatrix}
\frac{4}{3}k_-&\frac{i\sqrt{2}}{3}k_-&\frac{i\sqrt{2}}{\sqrt{3}}k_+\\
-\frac{i\sqrt{2}}{3}k_-&\frac{4i}{3}k_-&\frac{2i}{\sqrt{3}}k_+\\
\frac{i\sqrt{2}}{\sqrt{3}}k_+&\frac{2i}{\sqrt{3}}k_+&0
\end{bmatrix}
+Nk_z\begin{bmatrix}
0&-\frac{k_-}{\sqrt{2}}&-\frac{k_+}{\sqrt{6}}\\
-\frac{k_-}{\sqrt{2}}&0&\frac{k_+}{\sqrt{3}}\\
-\frac{k_+}{\sqrt{6}}&-\frac{k_+}{\sqrt{3}}&0
\end{bmatrix}
\end{eqnarray}
in which $k_\pm=k_x\pm ik_y$, and 
$
a=\frac{1}{3}(L+2M)(k_x^2+k_y^2+k_z^2)+2\xi_{SO}+\frac{\xi_T}{3},
b=\frac{1}{6}(L+5M)(k_x^2+k_y^2)+\frac{1}{3}(2L+M)k_z^2-\xi_{SO}+\frac{2\xi_T}{3},
c=\frac{1}{2}(L+M)(k_x^2+k_y^2)+Mk_z^2-\xi_{SO},
q=\frac{1}{3\sqrt{2}}(L-M)(k_x^2+k_y^2-2k_z^2)-\frac{\sqrt{2}}{3}\xi_T$, and 
$r=\frac{1}{\sqrt{6}}(L-M)(k_x^2-k_y^2)-\frac{2i}{\sqrt{6}}Nk_xk_y.
$
\\
\\

{\it Schrodinger equations-} The $\Gamma-$point Hamiltonian is given by
\begin{eqnarray}
H_0=H_\mathrm{full}(k_x=k_y=0).
\end{eqnarray}
Following the decomposition of the above Hamiltonian, the six eigenstates can be written in general forms as
\begin{eqnarray}
|\psi_\pm^\sigma\rangle&&= f_1^\pm(z)|\frac{1}{2},\sigma\frac{1}{2}\rangle+\sigma f_2^\pm(z)  |\frac{3}{2},\sigma\frac{1}{2}\rangle,\\
|\psi_0^\sigma\rangle&&=f_0(z)|\frac{3}{2},\sigma\frac{3}{2}\rangle,
\end{eqnarray}
corresponding to eigen-energies $E_\pm,E_0$, respectively, in which $\sigma=\pm$ stand for $\urow,\drow$, respectively. In the above, the spatial functions $(f_1^-,f_2^-)^T$ and $(f_1^+,f_2^+)^T$ are determined by solving the following coupled differential equations
\begin{eqnarray}\label{set1}
\frac{1}{3}
\begin{bmatrix}
-(L+2M)\partial_z^2+(6\xi_{SO}+\xi_T)&\sqrt{2}((L-M)\partial_z^2-\xi_T)\\
\sqrt{2}((L-M)\partial_z^2-\xi_T)&-(2L+M)\partial_z^2-(3\xi_{SO}-2\xi_T)
\end{bmatrix}\begin{pmatrix}
f_1^\pm(z)\\
f_2^\pm(z)
\end{pmatrix}+V(z)\begin{pmatrix}
f_1^\pm(z)\\
f_2^\pm(z)
\end{pmatrix}=E_\pm\begin{pmatrix}
f_1^\pm(z)\\
f_2^\pm(z)
\end{pmatrix},
\end{eqnarray}
and $f_0(z)$ is the solution of following differential equation
\begin{equation}\label{set2}
-Mf_0''(z)+V(z)f_0(z)=(E_0-\xi_{SO})f_0(z).
\end{equation}

{\it Boundary conditions -} In solving the above Schrodinger equations in a STO slab with finite thickness $d$, we assume that the electron is confined in a quantum well with open boundaries, i.e., $\psi(z=0,d)=0$. This assumption may be understood by considering the character of the boundaries.  On the one hand, at the LAO/STO interface, the La $t_{2g}$ states in LAO layer lie far above the Ti $t_{2g}$ states in the STO layer \cite{2DEG3}. Therefore,  the penetration of the Ti $t_{2g}$ states into the LAO can be neglected \cite{2DEG3} and one can assume an infinite barrier at the interface for the simplicity. On the other hand, the other surface of the STO slab is open to vacuum, and thus an infinite barrier potential is imposed. Similar boundary conditions are also applied for the $\delta$-doped STO slab, where both STO surfaces are open to vacuum.

{\it Solutions in zero field} - First, we find the the solutions of (\ref{set1}) and (\ref{set2}) in the absence of the electric field, which can be found as
\begin{eqnarray}
|\psi_\pm^{(0)},\sigma\rangle&&=f^{(0)}_\pm(z) \left(\cos{\chi_\pm}|\frac{1}{2},\sigma\frac{1}{2}\rangle+\sigma \sin{\chi_\pm}  |\frac{3}{2},\sigma\frac{1}{2}\rangle\right),\label{spinor}\\
|\psi_0^{(0)},\sigma\rangle&&=f^{(0)}_0(z) |\frac{3}{2},\sigma\frac{3}{2}\rangle,
\end{eqnarray}
 corresponding to eigen-energies
\begin{eqnarray}
E_\pm^{(0)}&&=\frac{1}{2}\left(\xi_{SO}+\xi_T+(M+L)\lambda^2\right)
\pm\frac{1}{2}\left[9\xi_{SO}^2-2\xi_{SO}(\xi_T+(L-M)\lambda^2)+(\xi_T+(L-M)\lambda^2)^2\right]^{1/2},\\
E_0^{(0)}&&=M\lambda^2-\xi_{SO},
\end{eqnarray}
respectively.
In the above, the spatial coefficients are simple sinusoidal functions as
\begin{eqnarray}
f^{(0)}_\pm(z) &&=f^{(0)}_0(z) =\sqrt{\frac{2}{d}}\sin{\lambda z},\ \ \lambda=n\frac{\pi}{d},
\end{eqnarray}
and the spinor angles in Eq. (\ref{spinor}) are read as
\begin{equation}\label{chi}
{\chi_\pm} =\tan^{-1}\left[\frac{2\xi_{SO}+\xi_T+\lambda^2(L+2M)-3E_\pm^{(0)}}{\sqrt{2}(\xi_T+(L-M)\lambda^2)}\right].
\end{equation}
It is easy to verify that $\chi=\chi_+=\chi_-+\frac{\pi}{2}$, which secures the orthogonality of the eigen-wavefunctions $\langle\psi_\tau^{(0)},\sigma |\psi_{\tau'}^{(0)},\sigma'\rangle=\delta_{\tau\tau'}\delta_{\sigma\sigma'}$. Furthermore, this angle is in the range $0<\chi<\arctan{\frac{1}{\sqrt{2}}}$.

{\it Solutions under applied field} - In the presence of the confinement potential $V(z)=\mathcal{E}_z z$, with $\mathcal{E}_z$ being the electric field, the solutions can be found via variational method \cite{Varmethod}, in which the eigen wavefunctions are expressed as
\begin{eqnarray}\label{ansatz}
|\psi_\pm^\sigma\rangle&&=f_\pm(z) \left(\cos{\chi_\pm}|\frac{1}{2},\sigma\frac{1}{2}\rangle+\sigma \sin{\chi_\pm}  |\frac{3}{2},\sigma\frac{1}{2}\rangle\right),\\
|\psi_0^\sigma\rangle&&=f_0(z) |\frac{3}{2},\sigma\frac{3}{2}\rangle,
\end{eqnarray}
in which 
\begin{eqnarray}
f_\tau(z)=C_\tau e^{-\beta_\tau  z/d} f_\tau^{(0)}(z),
\end{eqnarray}
where $C_\tau$ are the normalization constants given by
\begin{eqnarray}
\ \ C_\tau=\left[\frac{2\beta_\tau (\pi^2+\beta_\tau ^2)}{\pi^2(1-e^{-2\beta_\tau })}\right]^{1/2}.
\end{eqnarray}
 and $\beta_\tau $ are the variational parameters for minimization of the corresponding energies
\begin{eqnarray}
E_\tau&&=E_\tau ^{(0)}+\mathcal{E}_zd F(\beta_\tau),
\end{eqnarray}
in which
\begin{eqnarray}
F(\beta_\tau)=\frac{M_\tau \beta_\tau ^2}{\mathcal{E}_zd^3}+\frac{1}{2}\left(1+\frac{1}{\beta_\tau }+\frac{2\beta_\tau }{\pi^2+\beta_\tau ^2}-\coth{\beta_\tau }\right),
\end{eqnarray}
where $M_0=M,$ and $ M_\pm=\frac{1}{3}(L+2M)\cos^2{\chi_\pm}+\frac{1}{3}(2L+M)\sin^2{\chi_\pm}+\frac{\sqrt{2}}{3}(M-L) \sin{2\chi_\pm}$.

\begin{figure}[h]
  \includegraphics[width=\linewidth]{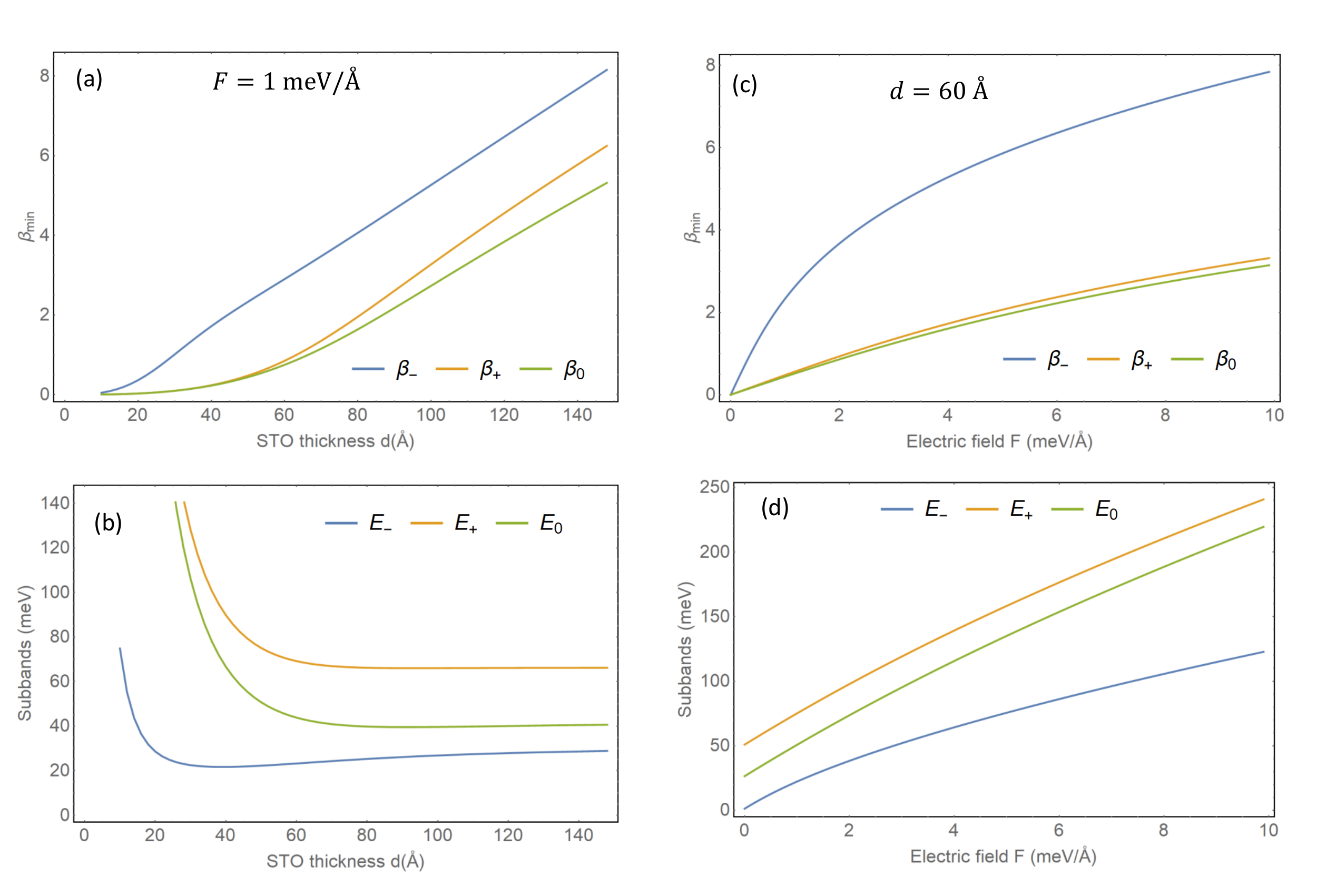}
  \caption{The variational parameters and subbands energies as functions STO thickness when $\mathcal{E}_z=5\ \mathrm{ meV/\AA}$ (a)-(b), and as functions of electric field when $d=60 \ \mathrm{\AA}$ (c)-(d), respectively.}
  \label{FigS4}
\end{figure}

The variational parameters can be obtained by minimizing the above energies, which increase as the electric field and the well width increase, as depicted in Fig. \ref{FigS4}. In the limit of weak electrostatic energy $\delta_\tau=\frac{\pi^2 \mathcal{E}_zd}{M_\tau\lambda^2}\ll 1$, the variational parameter and corresponding energy can be approximated as
\begin{eqnarray}\label{weak}
\beta_\tau=\frac{\pi^2-6}{12\pi^2}\delta_\tau, \ \ E_\tau=E_\tau^{(0)}+\frac{\mathcal{E}_zd}{2}-(M_\tau\lambda^2)\frac{(\pi^2-6)^2}{144\pi^2}\delta_\tau^2.
\end{eqnarray}
On the other hand, in the strong electrostatic energy limit $\delta\gg 1$, we have
\begin{eqnarray}\label{strong}
\beta_\tau=\frac{3^{1/3}\delta_\tau^{1/3}}{2^{2/3}},\ \ E_\tau=E_\tau^{(0)}+(M_\tau\lambda^2)\frac{3}{4}(6\delta_\tau)^{2/3}.
\end{eqnarray}

\section{General effective Hamiltonian}\label{S3}

Having derived the general expression of the eigenvectors at the $\Gamma$-point, 
we rearrange their sequence as 
\begin{eqnarray}
\{\psi_+^{\urow},\psi_+^{\drow},\psi_0^{\urow},\psi_0^{\drow},\psi_-^{\urow}, \psi_-^{\drow}\}.
\end{eqnarray}
 Map the full Hamiltonian (\ref{Hfull}) to the spin-orbital space of these vectors as $\mathcal{H}=\int_0^d dz(\psi_{\tau}^\sigma)^* H\psi_{\tau'}^{\sigma'}$, we obtain an effective Hamiltonian beyond the $\Gamma$-point, which can be written in block form as

\begin{eqnarray}\label{Hgen}
\mathcal{H}=
\begin{bmatrix}
\mathcal{H}_{+}&\mathcal{G} _{3}&\mathcal{G} _{1}\\
\mathcal{G} _{3}^\dag &\mathcal{H}_{0}&\mathcal{G} _{2}\\
\mathcal{G} _{1}^\dag &\mathcal{G} _{2}^\dag &\mathcal{H}_{-}
\end{bmatrix},
\end{eqnarray}

\subsection{Diagonal blocks}
The diagonal blocks of (\ref{Hgen}) are given by (for $\tau=0,\pm$)
\begin{eqnarray}\label{hpm}
\mathcal{H}_{\tau}&&=E_\tau+ \frac{\hbar^2 k^2}{2m^*_\tau} +\alpha_\tau (k_y \sigma_x-k_x \sigma_y ).
\end{eqnarray}
{\it Rashba spin-orbit coupling}
\begin{eqnarray}
\alpha_0&&=0,\\
 \alpha_{\pm}&&=\pm \frac{{\bar \gamma}_\pm}{3}(4\cos{2\chi}-\sqrt{2}\sin{2\chi})=\pm \frac{4\xi_{SO}{\bar \gamma}_\pm}{\Delta _I},
\end{eqnarray}
where ${\bar \gamma}_\pm=\int{dz \gamma(z)|f_\pm|^2}$, and have substituted $\chi=\chi_+$ so that $(4\cos{2\chi}-\sqrt{2}\sin{2\chi})=4\xi_{SO}/\Delta_I$, with
\begin{eqnarray}
\Delta_I=E_+^{(0)}-E_-^{(0)}=\sqrt{\eta^2+8\xi_{SO}^2},
\end{eqnarray}
with $\eta=\xi_{SO}+(M-L)\lambda^2$.

{\it Effective mass}
\begin{eqnarray}
\frac{1}{m^*_{0}}&&=\frac{1}{\hbar^2}(L+M),\\
\frac{1}{m^*_\pm}&&=  \frac{1}{2\hbar^2}(L+3 M)\pm\frac{1}{6\hbar^2}(M-L) \left(2 \sqrt{2} \sin (2 \chi )+\cos (2 \chi )\right),\nonumber\\
&&= \frac{1}{m^*_0}+\frac{(M-L)}{2\hbar^2}\left(1\pm\frac{\eta}{\Delta_I}\right).
\end{eqnarray}

{\it Effective ISB parameter-} As the ISB value rapidly decays in the second and deeper layers, we assume that it is only finite in the interface layer and zero in the bulk, i.e., $\gamma(z)=\gamma_0\theta(a-z)$, with $a$ being the lattice constant. In this case, the effective ISB parameter is read as
\begin{eqnarray}
{\bar \gamma}_\pm=\gamma_0\frac{\left(\coth \left(\beta _\pm \right)+1\right)}{2 \pi ^2} \left(\pi ^2-e^{-2 x \beta _\pm } \left(2 \beta _\pm ^2 \sin ^2(\pi  x)+\pi  \beta _\pm  \sin (2 \pi 
   x)+\pi ^2\right)\right),\ \ x=\frac{a}{d}.
\end{eqnarray}
which can be obtained once the variational parameters are known. In the the weak field regime, by substituting Eq. (\ref{weak}) the above can be explicitly expressed as
\begin{eqnarray}
{\bar \gamma}_\pm/\gamma_0=\frac{2 \pi ^2 a^3}{3d^3}+\frac{\left(\pi ^2-6\right) a^3 \mathcal{E}_z  }{18 M_\pm},
\end{eqnarray}
while in the strong field limit, we have
\begin{eqnarray}
{\bar \gamma}_\pm/\gamma_0=\frac{ a^3}{M_\pm}  \mathcal{E}_z +2\pi^2\left(\frac{4}{3 M_\pm^2}\right)^{2/3}\frac{a^3}{d^2} \mathcal{E}_z ^{1/3}.
\end{eqnarray}

\subsection{Off-diagonal blocks }
The off-diagonal blocks of (\ref{Hgen}) are given by
\begin{eqnarray}
\mathcal{G} _1&&=\epsilon_1(k)+\delta_1\left(k_y \sigma _x-k_x \sigma _y\right),\\
	\mathcal{G} _2&&=\epsilon_2(k)+\delta_2\left(k_x \sigma _x-k_y \sigma _y\right),\\
\mathcal{G} _3&&=\epsilon_3(k)+\delta_3\left(k_x \sigma _x-k_y \sigma _y\right),
\end{eqnarray}
where the hybridization parameters are
\begin{eqnarray}
\delta_{1}&&=\frac{1}{6}   \left(3 \sqrt{2} N \kappa_{+-}+2 \gamma_{+-} 
   \left(4 \sin (2 \chi )+\sqrt{2} \cos (2 \chi )\right)\right),\\
\delta_{2}&&=-\frac{i}{2 \sqrt{3}}  \left(\sqrt{2} \cos (\chi )+2 \sin (\chi )\right) \left(2
   \gamma_{-0}-N\kappa_{-0}\right),\\
		\delta_{3}&&=-\frac{i}{2 \sqrt{3}}  \left(\sqrt{2} \cos (\chi )-2 \sin (\chi )\right) \left(2
   \gamma_{+0}-N\kappa_{+0}\right),
\end{eqnarray}
and
\begin{eqnarray}
\epsilon_{1}&&=\frac{\rho_{+-}}{12}   (M-L) \left(k_x^2+k_y^2\right) \left(2 \sqrt{2} \cos (2 \chi )-\sin (2 \chi )\right),\\
	\epsilon_2&&=\frac{\rho_{-0}}{{2 \sqrt{3}}} \left(\cos (\chi )-\sqrt{2} \sin (\chi )\right) \left((M-L) \sigma _z
   \left(k_x^2-k_y^2\right)+2 i N k_x k_y\right),\\
\epsilon_{3}&&=-\frac{\rho_{+0}}{2 \sqrt{3}} \left(\sin (\chi )+\sqrt{2} \cos (\chi )\right) \left((M-L) \sigma _z
   \left(k_x^2-k_y^2\right)+2 i N k_x k_y\right),
\end{eqnarray}
In the above, we have defined the integrals $\rho_{\tau\tau'}=\int{dzf_\tau f_{\tau'}},\kappa_{\tau\tau'}=\int{dzf_\tau\partial_zf_{\tau'}},\gamma_{\tau\tau'}=\int{dz\gamma(z)f_\tau f_{\tau'}}$. Explicitly, they are given as
\begin{eqnarray}
\rho_{\tau\tau'}&&=C_\tau C_{\tau'}\frac{4(1-e^{-\beta_{\tau\tau'}})\pi^2}{\beta_{\tau\tau'}(4\pi^2+\beta_{\tau\tau'}^2)},\ \ \beta_{\tau\tau'}=\beta_\tau +\beta_{\tau'} \\
\kappa_{\tau\tau'}&&=\rho_{\tau\tau'}\frac{\beta_\tau -\beta_{\tau'} }{2d},\\
\gamma_{\tau\tau'}&&\approx \rho_{\tau\tau'}\frac{\beta_{\tau\tau'}(4\pi^2+\beta_{\tau\tau'}^2)}{5(1-e^{-\beta_{\tau\tau'}})}x^3,\ \ x=\frac{a}{d}\ll 1.
\end{eqnarray}

\section{Quasi-degenerate perturbation theory}\label{S4}

The Hamittonian (\ref{Hgen}) can be decomposed into three subspaces via the Lowdin partitioning \cite{winkler2003spin}, i.e.,
\begin{eqnarray}
\mathcal{H}=\begin{bmatrix}
\mathcal{H}_{+}&\mathcal{G} _{3}&\mathcal{G} _{1}\\
\mathcal{G} _{3}^\dag &\mathcal{H}_{0}&\mathcal{G} _{2}\\
\mathcal{G} _{1}^\dag &\mathcal{G} _{2}^\dag &\mathcal{H}_{-}
\end{bmatrix}\rightarrow
\begin{bmatrix}
\mathcal{H}_{le^+}&0&0\\
0&\mathcal{H}_{he}&0\\
0&0&\mathcal{H}_{le^-}
\end{bmatrix},
\end{eqnarray}
where the effective Hamiltonian of the each  subspace is derived as
\begin{eqnarray}
\mathcal{H}_{he}=\mathcal{H}_0+\frac{\mathcal{G} _2\cdot \mathcal{G} _2^\dag}{E_0-E_-}+\frac{\mathcal{G} _3^\dag\cdot \mathcal{G} _3}{E_0-E_+},\\
\mathcal{H}_{le^+}=\mathcal{H}_++\frac{\mathcal{G} _1\cdot \mathcal{G} _1^\dag}{E_+-E_-}+\frac{\mathcal{G} _3\cdot \mathcal{G} _3^\dag}{E_+-E_0},\\
\mathcal{H}_{le^-}=\mathcal{H}_-+\frac{\mathcal{G} _1^\dag\cdot \mathcal{G} _1 }{E_--E_+}+\frac{\mathcal{G} _2^\dag\cdot \mathcal{G} _2}{E_--E_0}.
\end{eqnarray}
Substituting the off-diagonal matrix $\mathcal{G}_i$ into the above, and to the leading order in the $\xi_{SO}$, we obtain the $k$-cubed corrections as
\begin{eqnarray}
\delta\mathcal{H}_{\tau}&&=\beta_3^{\tau}(k_x^2-k_y^2)(k_y\sigma_x-k_x\sigma_y)+\eta_3^{\tau}k_xk_y(k_x\sigma_x-k_y\sigma_y),
\end{eqnarray}
for $\tau=he,le^\pm$, in which the cubic SOC parameters are derived as
\begin{eqnarray}
&&\beta_3^{he}=\beta_3^{le^+}+\beta_3^{le^-},\ \eta_2^{he}=-(\eta_3^{le^+}+\eta_2^{le^-}),\label{a3}\\
&&\beta_3^{le+}=\frac{(M-L)\xi_{SO}}{\Delta_I}\frac{R_+}{\Delta_+},\ \ \eta_3^{le+}=\frac{2N\xi_{SO}}{\Delta_I}\frac{R_+}{\Delta_+},\\
&&\beta_3^{le-}=\frac{(M-L)\xi_{SO}}{\Delta_I}\frac{R_-}{\Delta_-},\ \ \eta_3^{le-}=\frac{2N\xi_{SO}}{\Delta_I}\frac{R_-}{\Delta_-},
\end{eqnarray}
in which $R_+=(2\gamma_{+0}-N\kappa_{+0})\rho_{+0},\ \ R_-=(2\gamma_{-0}-N\kappa_{-0})\rho_{-0},\ \
\Delta_+={E_+-E_0},\ \
\Delta_-={(E_0-E_-)}$. As the $le^+$ band lie far above the $he$ and $le^-$ bands, its corresponding wavefunction is more spreading into the bulk. As a consequence, one can show that $\gamma_{+0}\ll\gamma_{-0}, \rho_{+0}\ll\rho_{-0}$, and $\kappa_{+0}\ll\kappa_{-0}$, so that $P_+\ll P_-$. Therefore, in Eqs.\ref{a3} we retain only the second term,i.e.,  $\beta_3^{he}=\beta_3^{le^-},\ \eta_2^{he}=-eta_2^{le^-}$.

In the weak electric field regime, we have following approximations to the leading order in the E-field
\begin{eqnarray}
R_\pm&&\approx\frac{4\pi^2\gamma_0a^3}{3d^3}+\mathcal{E}_z\left[\frac{\gamma_0a^3(\pi^2-6)}{18}\left(\frac{1}{M_0}+\frac{1}{M_\pm}\right)+\frac{Nd^2(\pi^2-6)}{24\pi^2}\left(\frac{1}{M_\pm}-\frac{1}{M_0}\right)\right]\\
 \Delta_\pm&&\approx \Delta_\pm^{(0)}\pm \mathcal{E}_z^2\frac{(\pi^2-6)^2d^4}{144\pi^4}\left(\frac{1}{M_0}-\frac{1}{M_\pm}\right), 
\end{eqnarray}
where $\Delta_\pm^{(0)}=\pm(E_\pm^{(0)}-E_0^{(0)})$. The above equations show the dependence of the cubic RSOC on the electric field as $\sim \frac{R_\pm}{\Delta_\pm}\sim\frac{1}{\mathcal{E}_z}$.

\section{Orbital occupancy} \label{S5}

The occupancy of the orbitals in each band can be obtained by representing the eigenstates of the effective Hamiltonian in terms of the $t_{2g}$ basis as
\begin{eqnarray}
|\Psi_\tau^\sigma\rangle=\sum_i c_i^\tau |i\rangle
\end{eqnarray}
for $\tau=0,\pm$, and $i=xy,yz,zx$. Then $\mathcal{N}_i^\tau=|c_i^\tau|^2$ is referred to as the occupancy of the $i$-orbital in the $\tau$ bands. Around the $\Gamma$-point, the occupancies can be analytically obtained as
\begin{eqnarray}
\mathcal{N}_{xy}^{he}=0,\ \ && \mathcal{N}_{yz}^{he}=\mathcal{N}_{zx}^{he}=\frac{1}{2},\\
\mathcal{N}_{xy}^{le^-}=\frac{1}{3}(\sqrt{2}\cos\chi+\sin\chi)^2,\ \ && \mathcal{N}_{yz}^{le^-}=\mathcal{N}_{zx}^{le^-}=\frac{1}{6}(\cos{\chi}-\sqrt{2}\sin{\chi})^2,\\
\mathcal{N}_{xy}^{le^+}=\frac{1}{3}(\cos\chi-\sqrt{2}\sin\chi)^2,\ \ && \mathcal{N}_{yz}^{le^+}=\mathcal{N}_{zx}^{le^+}=\frac{1}{6}(\sqrt{2}\cos{\chi}+\sin{\chi})^2.
\end{eqnarray}
where $\tau=he,le^\pm$ are equivalent to $0,\pm$ bands, respectively. It can be verified that the orbital occupancies satisfy $\sum_i \mathcal{N}_i^\tau=1$ for each band.

Substitute $\chi=\chi_+$ given in (\ref{chi}), the above expressions reduce to
\begin{eqnarray}
\mathcal{N}_{xy}^{he}=0,\ \ \mathcal{N}_{xy}^{le^\pm}&&=\frac{1}{2}\left(1\mp\frac{\eta}{\Delta_I}\right),\\
\mathcal{N}_{yz/zx}^\tau&&=\frac{1}{2}(1-\mathcal{N}_{xy}^\tau).
\end{eqnarray}

\end{widetext}

\end{document}